\documentclass[journal]{IEEEtran}
\ifCLASSINFOpdf
\else
\fi

\usepackage[hyphens]{url}
\usepackage[hidelinks]{hyperref}
\hypersetup{breaklinks=true}
\usepackage{verbatim}
\usepackage{multirow}
\usepackage{cite}
\usepackage{amsmath,amssymb,amsfonts}
\usepackage[table]{xcolor}

\usepackage[linesnumbered,ruled,vlined]{algorithm2e}
\usepackage{algorithmic}
\usepackage{graphicx}
\DeclareGraphicsExtensions{.pdf,.jpeg,.png}
\usepackage{stfloats}
\usepackage{textcomp}
\usepackage{xcolor}
\usepackage{multicol}
\usepackage{multirow}
\usepackage{afterpage}
\usepackage{siunitx}
\usepackage{nccmath}
\usepackage{array}
\usepackage{makecell}
\usepackage{hhline}
\usepackage{booktabs}
    
\newcommand{\thickhline}{
    \noalign {\ifnum 0=`}\fi \hrule height 1pt
    \futurelet \reserved@a \@xhline
}

\makeatletter
\newcommand{\vast}{\bBigg@{4}}
\newcommand{\Vast}{\bBigg@{5}}
\makeatother

\begin{document}

\title{Zonotope Shadow and Reflection Matching: A Novel GNSS Reflection-Based Framework for Enhanced Positioning Accuracy in Urban Areas}

\author{Sanghyun~Kim and Jiwon~Seo,~\IEEEmembership{Member,~IEEE}
\thanks{Manuscript received February 00, 2025; }
\thanks{This work was supported in part by the National Research Foundation of Korea (NRF), funded by the Korean government (Ministry of Science and ICT, MSIT), under Grant RS-2024-00358298; 
in part by the Unmanned Vehicles Core Technology Research and Development Program through the NRF and the Unmanned Vehicle Advanced Research Center (UVARC), funded by the MSIT, Republic of Korea, under Grant RS-2020-NR046546;
in part by the MSIT, Korea, under the Information Technology Research Center (ITRC) support program, supervised by the Institute of Information \& Communications Technology Planning \& Evaluation (IITP), under Grant IITP-2024-RS-2024-00437494;
in part by Grant RS-2024-00407003 from the ``Development of Advanced Technology for Terrestrial Radionavigation System'' project, funded by the Ministry of Oceans and Fisheries, Republic of Korea;
and in part by the Korea Aerospace Administration (KASA), under Grant RS-2022-NR067078.
\textit{(Corresponding Author: Jiwon Seo.)}}
\thanks{Sanghyun Kim is with the School of Integrated Technology, Yonsei University, Incheon 21983, Republic of Korea (e-mail: sanghyun.kim@yonsei.ac.kr).}
\thanks{Jiwon~Seo is with the School of Integrated Technology, Yonsei University, Incheon 21983, Republic of Korea and also with the Department of Convergence IT Engineering, Pohang University of Science and Technology, Pohang 37673, Republic of Korea (e-mail: \mbox{jiwon.seo@yonsei.ac.kr}).}
}

%
%

\markboth{}%
{Kim \MakeLowercase{\textit{and}} Seo: Zonotope Shadow and Reflection Matching for
Enhancing Positioning Accuracy in Urban Areas}
%



\maketitle


\begin{abstract} 
In urban areas, signal reception conditions are often poor due to reflections from buildings, resulting in inaccurate global navigation satellite system (GNSS)-based positioning. 
Various 3D-mapping-aided (3DMA) GNSS techniques, including shadow matching, have been proposed to address this issue. 
However, conventional shadow matching estimates positions in a discretized manner. 
The accuracy of this approach is limited by the resolution of the grid points representing the candidate receiver positions, making it difficult to achieve robust urban positioning and to ensure that the position estimate satisfies user-specified protection levels or safety bounds. 
To overcome these limitations, zonotope shadow matching (ZSM) has been proposed, which utilizes a set-based position estimate rather than grid-based estimates. 
ZSM calculates the GNSS shadow---an area on the ground where the line-of-sight (LOS) is blocked and only non-line-of-sight (NLOS) signals can be received---to estimate the receiver’s position set. 
ZSM distinguishes between LOS and NLOS satellites, determining that the receiver is inside the GNSS shadow if the satellite is NLOS and outside if the satellite is LOS. 
However, relying solely on GNSS shadows limits the ability to sufficiently reduce the size of the receiver position set and to precisely estimate the receiver’s location. 
To address this, we propose zonotope shadow and reflection matching (ZSRM) to enhance positioning accuracy in urban areas. 
ZSRM incorporates both GNSS shadows and GNSS reflections to estimate the receiver’s position. 
A GNSS reflection is defined as an area on the ground where both LOS and NLOS signals are received simultaneously. 
Unlike ZSM, ZSRM classifies satellites into three types: LOS-only, LOS + NLOS, and NLOS-only. 
Based on this classification, the receiver’s position is determined in one of three ways: inside the GNSS shadow (if NLOS-only), inside the GNSS reflection (if LOS + NLOS), or outside both areas (if LOS-only). 
Using this framework, ZSRM can estimate the receiver’s position more precisely than ZSM. 
The proposed ZSRM technique is validated through field tests using GNSS signals collected in an urban environment. 
Consequently, the RMS horizontal position error of ZSRM improved by 10.0\% to 53.6\% compared with ZSM, while the RMS cross-street and along-street position bounds improved by 18.0\% to 50.1\% and 30.7\% to 59.3\%, respectively. 
\end{abstract}

\begin{IEEEkeywords}
GNSS-based localization, 3D-mapping-aided GNSS, urban positioning, constrained zonotope, GNSS shadow, GNSS reflection, set-based positioning.
\end{IEEEkeywords}

%
\IEEEpeerreviewmaketitle

\section{Introduction}
\label{sec:Introduction}

\IEEEPARstart{I}{ntelligent} transportation systems, such as autonomous vehicles, heavily rely on the global navigation satellite system (GNSS) for accurate positioning services \cite{Lee22:Urban, Ma20:Articial}. 
However, GNSS-based positioning is often inaccurate in urban areas due to poor signal reception \cite{Kim22:Machine, Lee23:Seamless, Zhu18}. 
Signals can be blocked, reflected, or diffracted by buildings, which reduces satellite visibility and introduces measurement errors, potentially resulting in position errors exceeding 100~m \cite{MacGougan02}.

\begin{figure}
    \centering
    \includegraphics[width=0.7\linewidth]{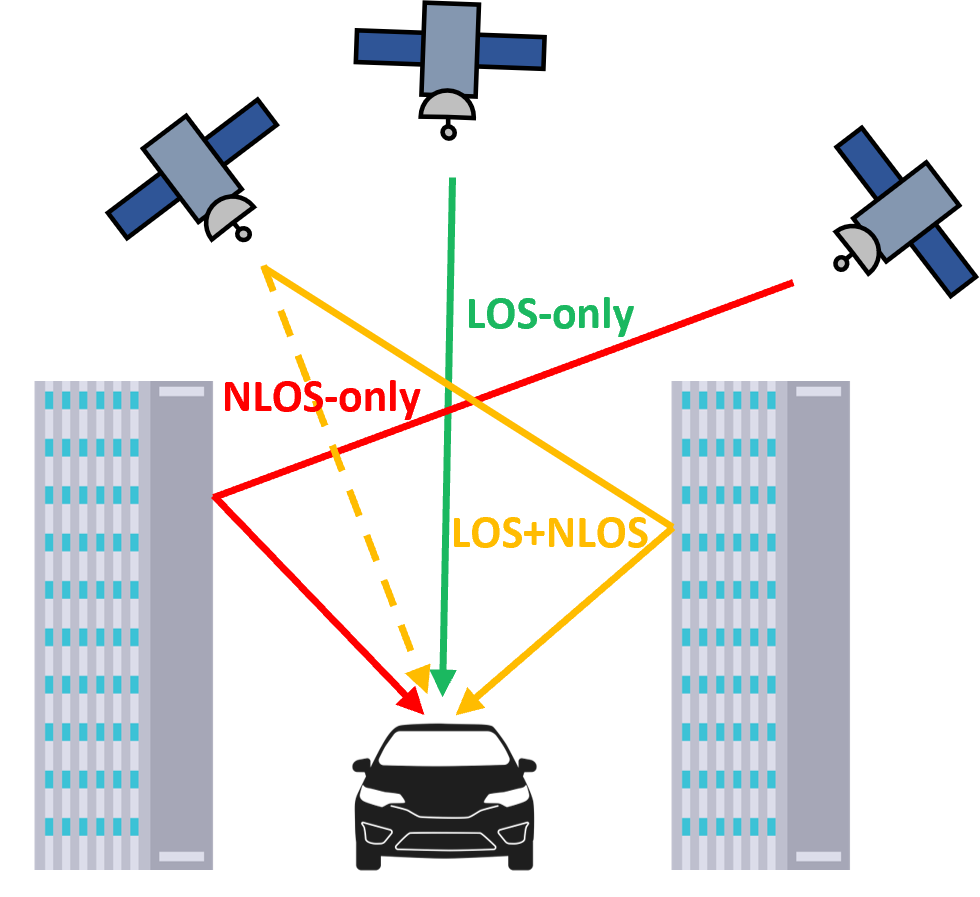}
    \caption{Illustration of GNSS signal reception conditions: LOS-only, NLOS-only, and LOS + NLOS.}
    \label{fig:SRC}
\end{figure}

In general, GNSS signals in urban environments can be received under three distinct conditions, as shown in Fig.~\ref{fig:SRC} \cite{Adjrad18:Intelligent}. 
Line-of-sight-only (LOS-only) reception refers to a condition in which only a direct signal is received. 
LOS-only signals experience no propagation delays due to reflection and therefore, theoretically, no multipath range errors occur. 
Second, non-line-of-sight-only (NLOS-only) reception occurs when only reflected signals are received. 
NLOS-only signals exhibit positive range errors resulting from propagation delays caused by signal reflections. 
These errors are typically on the order of several tens of meters but can become significantly larger because they are unbounded. 
Third, LOS + NLOS reception occurs when both LOS and NLOS signals are received simultaneously. 
Unlike NLOS-only signals, LOS + NLOS signals can result in both positive and negative range errors. 
Additionally, the range errors of LOS + NLOS signals are bounded because the correlation process suppresses reflected signals with delays exceeding 1.5 chips (e.g., one GPS L1 C/A code chip corresponds to approximately 300~m) \cite{Enge1994:global}. 
Although NLOS-only and LOS + NLOS signals are often grouped together as multipath, it is crucial to recognize that they are distinct phenomena requiring different mitigation techniques \cite{Groves13:GNSS}.

\begin{figure}
    \centering
    \includegraphics[width=0.85\linewidth]{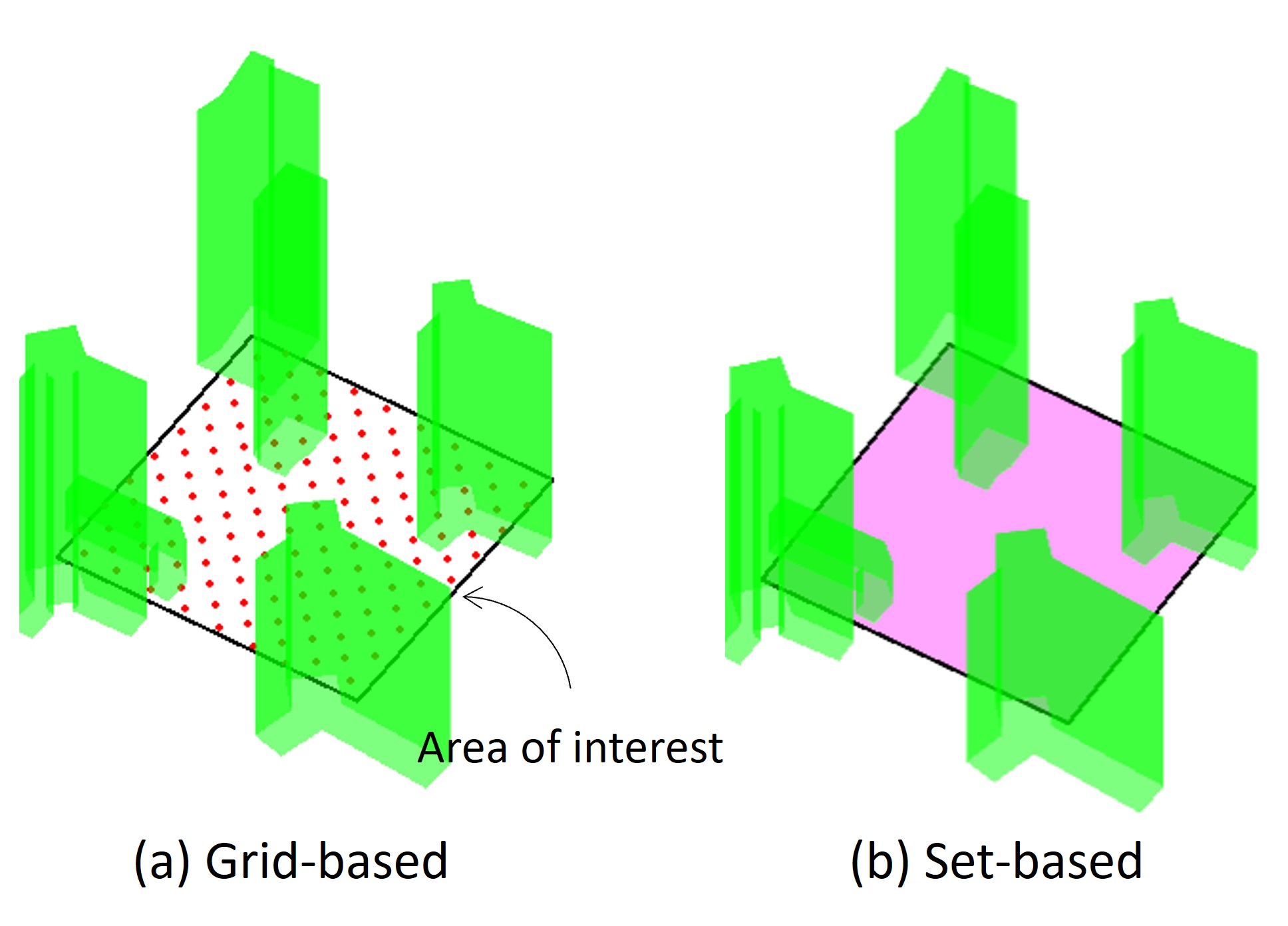}
    \caption{(a) In the grid-based approach, each red grid point represents a candidate receiver position. (b) In the set-based approach, the magenta region represents the entire set of candidate receiver positions.}
    \label{fig:GridvsSet}
\end{figure}

\begin{figure*}
    \centering
    \includegraphics[width=1.00\linewidth]{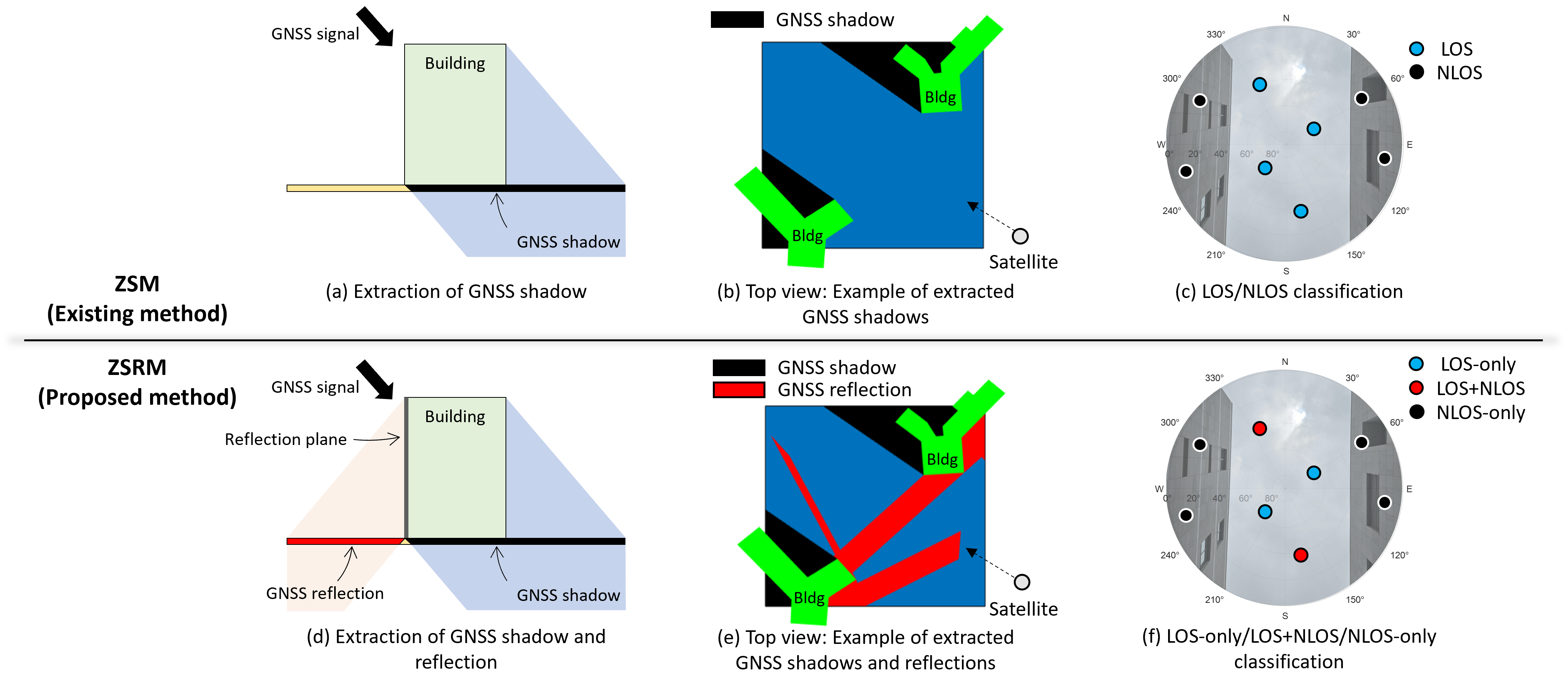}
    \caption{Comparison of key differences between the existing ZSM and the proposed ZSRM methods. Subfigures (a)--(c) illustrate ZSM, while subfigures (d)--(f) illustrate ZSRM.}
    \label{fig:ZSMvsZSRM}
\end{figure*}

Various methods have been developed to address the challenge of GNSS positioning in urban areas, including measurement weighting \cite{Lesouple18:Multipath}, consistency checking \cite{Zhang18}, terrain height aiding \cite{Adjrad17:Enhancing}, and GNSS/inertial navigation system (INS) integration \cite{Zhao24}. 
In addition to these traditional methods, an innovative solution involves the use of three-dimensional (3D) city models, commonly referred to as 3D-mapping-aided (3DMA) GNSS \cite{Zhong22:Multi, Lee23:Nonlinear}. 
In particular, with the increasing availability of high-precision 3D city models and advances in computing capabilities, 3DMA GNSS has the potential to significantly enhance positioning performance in urban areas \cite{Miura15:GPS}.

From a methodological standpoint, 3DMA GNSS can be categorized into two primary types: ray tracing and shadow matching. 
Ray tracing calculates all possible signal propagation paths between the satellite and the receiver, enabling the correction of pseudorange biases in the signal \cite{Suzuki12:GNSS}. 
However, the significant computational cost associated with ray tracing makes it impractical for real-time applications \cite{vanDiggelen21:End, Lee24:Efficient}.  
Another major category of 3DMA GNSS is shadow matching. 
Shadow matching begins by overlaying a grid across an area of interest at a uniform resolution, with each grid point representing a candidate receiver position \cite{Groves11:Shadow}. 
Subsequently, the visibility of each satellite is calculated and evaluated for every grid point, which can be rapidly obtained using a precomputed building boundary lookup table. 
The building boundary represents the minimum elevation angle required to receive satellite signals at specific azimuths, and can be derived from a 3D city model. 
The lookup table method significantly reduces the computational load, making shadow matching a practical approach for enhancing urban GNSS accuracy in real time \cite{Wang13:GNSS}.

However, challenges remain in shadow matching \cite{Groves15:GNSS}. 
One of the main issues is that shadow matching inherently estimates positions in a discretized manner. 
In this method, positioning accuracy varies depending on the grid resolution. 
While finer discretization can improve accuracy, it also increases computational costs, which may negatively impact real-time applications. 
Furthermore, the grid-based approach makes it difficult to satisfy user-defined protection levels or safety constraints \cite{Bhamidipati22:Set}.

To address the limitations associated with the grid-based approach, a set-based shadow matching approach was proposed. 
As shown in Fig.~\ref{fig:GridvsSet}, the grid-based approach represents the receiver position using predefined discrete grid points, while the set-based approach defines the receiver position and surrounding objects as geometric sets. 
In the set-based method, once the final set area is determined by the position estimation algorithm, the centroid of the area is used as the receiver position \cite{Bhamidipati22:Set}. 
Bhamidipati \textit{et al.} \cite{Bhamidipati22:Set} proposed the zonotope shadow matching (ZSM) algorithm, which utilizes zonotopes to comprehensively address shadow matching with set-based objects.

A zonotope is a convex, symmetrical polytope primarily used to represent sets in various control algorithms, such as path planning and collision avoidance \cite{Althoff10:Computing, Kousik19:Safe}. 
Zonotopes enable rapid computation in situations requiring numerous set operations. 
Additionally, zonotopes demonstrate scalability in set representations by expanding to a constrained zonotope, which allows the representation of any convex polytope without symmetry limitations \cite{Scott16:Constrained}.

ZSM \cite{Bhamidipati22:Set, Kim25:Set} uses constrained zonotopes to represent each building and the ground plane in a 3D city model. 
It then generates GNSS shadows for each satellite–building pair. 
These GNSS shadows represent two-dimensional (2D) areas on the ground plane where LOS signals are blocked. 
An example of a GNSS shadow for a specific satellite is shown in Fig.~\ref{fig:ZSMvsZSRM}(a), and a top view of the shadow is shown in Fig.~\ref{fig:ZSMvsZSRM}(b). 
After computing the GNSS shadow, the receiver position is estimated based on the signal reception condition of the satellite. 
For instance, if the satellite is LOS (i.e., LOS-only or LOS + NLOS), the area outside the GNSS shadow (e.g., the blue area in Fig.~\ref{fig:ZSMvsZSRM}(b)) is considered the receiver position set. 
Conversely, if the satellite is NLOS (i.e., NLOS-only), the area inside the GNSS shadow (e.g., the black area in Fig.~\ref{fig:ZSMvsZSRM}(b)) is taken as the receiver position set.

The satellite signal reception condition can be classified using various methods, including statistical techniques based on the carrier-to-noise-density ratio ($C/N_0$) \cite{Kim21:GPS}, as well as machine-learning (ML) and deep-learning (DL) approaches \cite{Hsu17, Zeng24} that leverage various GNSS measurements as features. 
Finally, the receiver’s position set is gradually refined by iteratively applying this position-determination algorithm to all visible satellites.

Although ZSM has improved positioning accuracy in urban environments, it still has limitations. 
As shown in Fig.~\ref{fig:ZSMvsZSRM}(b), the area outside the GNSS shadow is often significantly larger than the area inside. 
Therefore, in environments where LOS (i.e., LOS-only or LOS + NLOS) satellites dominate, ZSM may struggle to effectively reduce the size of the receiver position set. 
This issue becomes particularly problematic in environments where multiple LOS + NLOS signals are received, making it difficult to achieve reliable positioning results. 
Additionally, GNSS shadows typically occur in the cross-street direction, which can improve positioning performance in that direction but lead to relatively lower performance in the along-street direction. 
This limitation applies not only to ZSM, but also to conventional shadow matching methods. 
Therefore, a novel GNSS localization method capable of more precisely estimating the receiver position is necessary.

In this paper, we present an improved positioning technique that addresses the limitations of ZSM. 
Specifically, we introduce zonotope shadow and reflection matching (ZSRM), a method that estimates set-based receiver positions more precisely by incorporating GNSS reflections in addition to GNSS shadows. 
ZSRM computes both GNSS shadows and reflections for each satellite--building pair using a 3D city model represented as constrained zonotopes.

This study introduces the concept of GNSS reflection for the first time in the context of 3DMA GNSS localization. 
A GNSS reflection represents a 2D area on the ground plane where both LOS and NLOS signals from a satellite are received simultaneously. 
As shown in Fig.~\ref{fig:ZSMvsZSRM}(d), the GNSS reflection is distinctly different from the GNSS shadow. 
By incorporating GNSS reflections, the estimation algorithm for the receiver’s position can be further refined, thereby improving positioning performance. 
In ZSM, the receiver’s position is determined outside the GNSS shadow (e.g., the blue area in Fig.~\ref{fig:ZSMvsZSRM}(b)) for both LOS-only and LOS + NLOS satellites. 
In contrast, with ZSRM, the receiver’s position is determined outside both the GNSS shadow and reflection (e.g., the blue area in Fig.~\ref{fig:ZSMvsZSRM}(e)) for LOS-only satellites, whereas it is determined within the GNSS reflection area (e.g., the red area in Fig.~\ref{fig:ZSMvsZSRM}(e)) for LOS + NLOS satellites. 
To enable this process, we further classified LOS satellites into LOS-only and LOS + NLOS categories, as shown in Figs.~\ref{fig:ZSMvsZSRM}(c) and \ref{fig:ZSMvsZSRM}(f).

Additionally, as shown in Fig.~\ref{fig:ZSMvsZSRM}(e), GNSS reflections occur in multiple directions owing to signal reflections from various building planes at different angles. 
Therefore, if these GNSS reflections are computed and incorporated into a positioning algorithm, they can potentially improve accuracy in both the along-street and cross-street directions.

Our key contributions are as follows: 
\begin{itemize}
  \item We introduce the novel concept of GNSS reflection and develop a method to compute it using a 3D city model represented by constrained zonotopes; 
  \item We propose the ZSRM algorithm, which estimates the set-based receiver position by incorporating both GNSS shadows and reflections; 
  \item We validate the positioning performance of ZSRM against that of ZSM using GPS and Galileo signals collected in an urban environment. 
\end{itemize}

The remainder of this paper is organized as follows. 
Section~\ref{sec:Set} introduces the set representation and operations using constrained zonotopes. 
Section~\ref{sec:Existing} presents the existing ZSM algorithm. 
Section~\ref{sec:Proposed} details the proposed ZSRM algorithm. 
Section~\ref{sec:Field} reports the results of field tests conducted in an urban environment to validate the ZSRM algorithm.  
Finally, Section~\ref{sec:Conclusion} concludes the paper.

\section{Constrained Zonotope Preliminaries}
\label{sec:Set}

The proposed ZSRM technique employs constrained zonotopes to represent set-based objects and efficiently compute GNSS shadows and reflections through set operations. 
In this section, we define a constrained zonotope and introduce three fundamental set operations associated with constrained zonotopes.

As previously explained, a zonotope is a convex, symmetrical polytope used in various algorithms to represent a set. 
Any zonotope can be expressed as a constrained zonotope, which generalizes zonotopes by removing their inherent symmetry constraints and can represent all convex polytopes. 
A constrained zonotope can be expressed as follows \cite{Scott16:Constrained}: 
\begin{equation}
\begin{split}
  Z & = \mathrm{zono}(\mathbf{c}, G, A, \mathbf{b}) \\
  & = \{ \mathbf{c}+G\beta \in \mathbb{R}^n \, \lvert \, \beta \in [-1,1]^m \, \textrm{and} \, A\beta=\mathbf{b} \} \subset \mathbb{R}^n
  \label{eqn:ConZono}
\end{split}
\end{equation}
where $\mathbf{c}\in\mathbb{R}^n$ is the center, $G=[g_1,g_2,\cdots,g_m]\in\mathbb{R}^{n \times m}$ is a generator matrix consisting of $m$ generators, and $A\in\mathbb{R}^{p \times m}$, $\mathbf{b}\in\mathbb{R}^p$ are parameters that define linear constraints. 
Each generator corresponds to a direction and magnitude that influence the shape of the zonotope. 
Notably, when the zonotope is unconstrained (i.e., $A$ and $\mathbf{b}$ are not defined), it reduces to a standard zonotope, denoted as $\mathrm{zono}(\mathbf{c},G,[\,],[\,])$.

\begin{figure}
    \centering
    \includegraphics[width=0.95\linewidth]{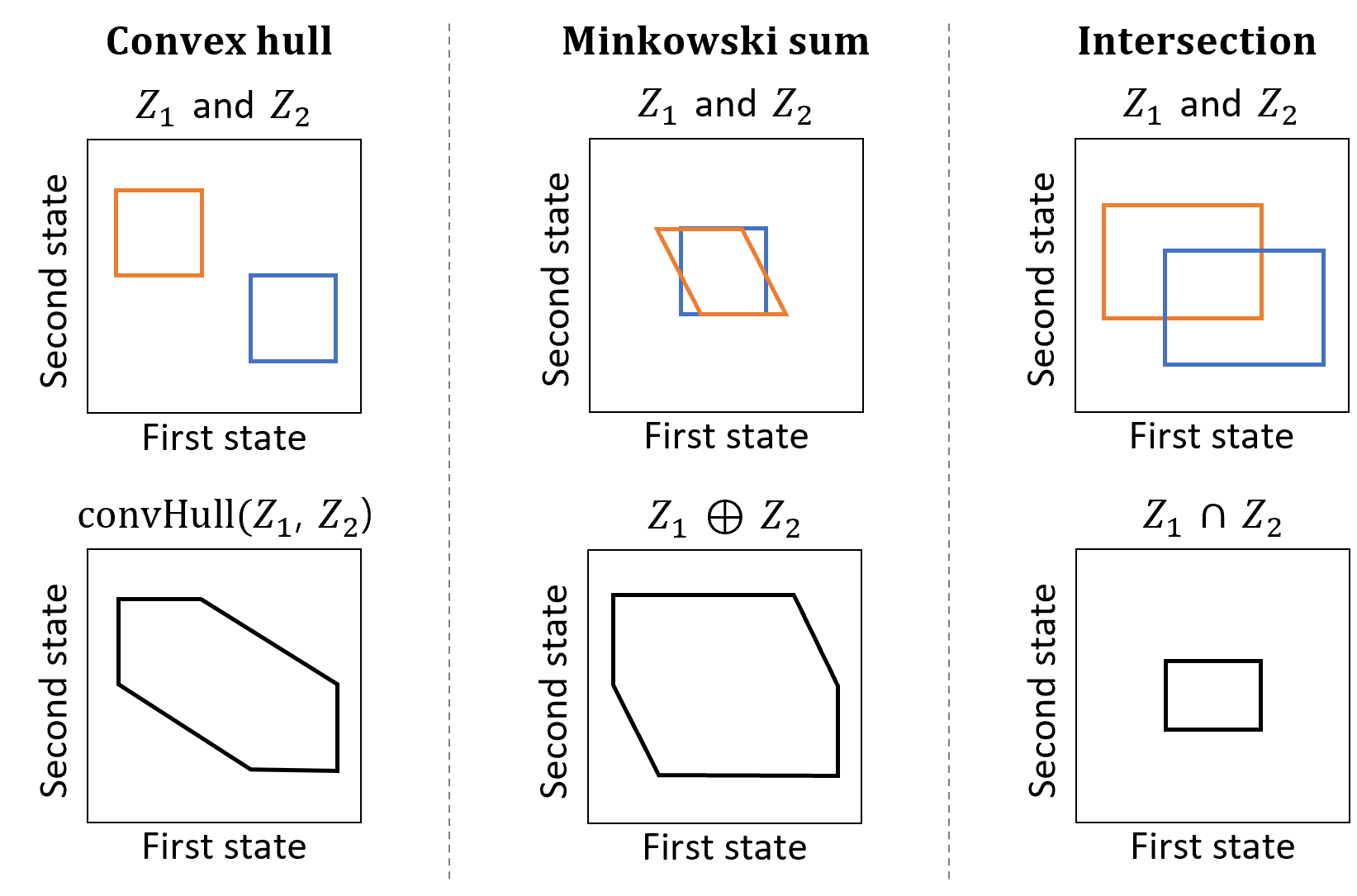}
    \caption{Examples of set operations (convex hull, Minkowski sum, and intersection) with 2D constrained zonotopes. The orange and blue 2D constrained zonotopes represent $Z_1$ and $Z_2$, respectively, while the black 2D constrained zonotopes are the results of the set operations between them (adapted from \cite{Althoff24:CORA}).}
    \label{fig:SetOperations}
\end{figure}

Constrained zonotopes exhibit closure under specific set operations, such as convex hull \cite{Raghuraman22:Set}, Minkowski sum \cite{Althoff14:Online}, and intersection \cite{Scott16:Constrained}. 
The results of these operations for two given 2D constrained zonotopes are illustrated in Fig.~\ref{fig:SetOperations}. 
Because constrained zonotopes are closed under these operations, the outcomes can also be represented as constrained zonotopes. 
In this study, we represent the 3D city model and satellite signal direction vectors as constrained zonotopes, and compute GNSS shadows and reflections using set operations between them.

\section{Existing Zonotope Shadow Matching Algorithm}
\label{sec:Existing}

In this section, we describe the ZSM algorithm proposed by Bhamidipati \textit{et al.} \cite{Bhamidipati22:Set}. 
ZSM represents the various components involved in shadow matching as constrained zonotopes and employs them to estimate a set-based receiver position. 
We first explain how a standard 3D city model is preprocessed into a zonotope-represented 3D city model. 
We then describe the computation of GNSS shadows and explain how the set-based receiver position is estimated.

\subsection{Preprocessing Standard 3D City Model}
\label{sec:Preprocessing}

ZSM employs a 3D city model composed of a ground plane and a collection of buildings. 
Standard 3D city models generated using computer-aided design (CAD) software often consist of triangles represented by vertex coordinates. 
Accordingly, the ground plane and buildings can be expressed as 
\begin{equation}
\label{eqn:3Dcitymodel}
\begin{split}
\mathcal{G} &= \{T_k^\mathrm{grnd}\}_{k=1}^{n_\mathrm{grnd}} \\
\mathcal{B} &= \{B_i\}_{i=1}^{n_\mathrm{bldg}}, \, \text{where} \, B_i = \bigcup\limits_{l=1}^{n_i}T_l^\mathrm{bldg}
\end{split}
\end{equation}
where $\mathcal{G}$ denotes the ground plane comprising a collection of triangles $T_k^\mathrm{grnd}$, and $\mathcal{B}$ denotes the set of all buildings $B_i$. 
Here, $n_\mathrm{grnd} \in \mathbb{N}$ is the number of triangles representing the ground plane, and $n_\mathrm{bldg} \in \mathbb{N}$ is the total number of buildings. 
Each $B_i$ is a union of triangles $T_l^\mathrm{bldg}$, with $n_i \in \mathbb{N}$ denoting the number of triangles that constitute $B_i$.

To convert a standard 3D city model into a zonotope-represented 3D city model, each triangle constituting the standard 3D city model is transformed into a constrained zonotope, as shown in the following equation: 
\begin{equation}
\label{eqn:Preprocessing}
\begin{split}
T &= \mathrm{convHull}(t_1,t_2,t_3) \\
&= \mathrm{convHull}(\mathrm{convHull}(t_1,t_2),t_3)
\end{split}
\end{equation}
where $T$ is a triangle consisting of vertices $\mathbf{t}_i \, (i=1,2,3) \in \mathbb{R}^3$, and each vertex can be represented as a constrained zonotope, $t_i = \mathrm{zono}(\mathbf{t}_i,[\,],[\,],[\,])$. 
A triangle can be expressed as the convex hull of its three vertices. 
Equivalently, this can be obtained by applying the convex hull between two vertices twice, as $\mathrm{convHull}(\mathrm{convHull}(t_1,t_2),t_3)$. 
The resulting triangle, denoted as $T$, can thus be represented as a constrained zonotope through successive convex hull operations.

Using (\ref{eqn:Preprocessing}), the vertex-represented triangles $T_k^\mathrm{grnd}$ and $T_l^\mathrm{bldg}$ in (\ref{eqn:3Dcitymodel}) can be transformed into constrained zonotopes, denoted as $G_k$ and $Z_l$, respectively. 
Consequently, a zonotope-represented 3D city model consisting of a ground plane and a set of buildings is obtained as follows: 
\begin{equation}
\label{eqn:3DcitymodelZono}
\begin{split}
\mathcal{G} &= \{G_k\}_{k=1}^{n_\mathrm{grnd}} \\
\mathcal{B} &= \{B_i\}_{i=1}^{n_\mathrm{bldg}}, \, \text{where} \, B_i = \bigcup\limits_{l=1}^{n_i}Z_l
\end{split}
\end{equation}

\subsection{Computing GNSS Shadows}
\label{sec:Shadows}

Here, we describe the process of computing GNSS shadows using constrained zonotopes. 
The computation procedure consists of three main steps: 1) computing the shadow directions, 2) computing the shadow volumes, and 3) computing the GNSS shadows.

\subsubsection{Computing Shadow Directions}
\label{sec:ShadowDirections}

The shadow direction is defined as the unit vector from the satellite to the building. 
It is assumed to be a single direction from the satellite to a representative point of the building. 
This assumption is reasonable because the distance from the satellite to the building is significantly larger than the building size \cite{Bhamidipati22:Set}. 
The shadow direction $\hat{\mathbf{l}}_{i,j}$ for the $i$-th building and the $j$-th satellite is computed as 
\begin{equation}
\label{eqn:ShadowDirection}
\hat{\mathbf{l}}_{i,j}=\frac{\mathbf{b}_j-\mathbf{s}_i}{\|\mathbf{b}_j-\mathbf{s}_i\|_2}
\end{equation}
where $\mathbf{s}_j \in \mathbb{R}^3$ denotes the position of the $j$-th satellite, and $\mathbf{b}_i \in \mathbb{R}^3$ denotes a representative point of building $B_i$.
This representative point is defined as the mean of the vertices of all constrained zonotopes that constitute $B_i$.

The method proposed by Althoff \cite{Althoff15:An} and Matt \cite{Matt21:Analyze} is used to obtain the vertices of $B_i$. 
The result of the method is summarized as follows: 
\begin{equation}
\label{eqn:Vertices}
  \{\mathbf{v}_i\}_{i=1}^{n_\mathrm{verts}} = \mathrm{vertices}(\mathcal{Z}), \, \text{where} \, \mathcal{Z} = \{Z_i\}_{i=1}^{n_z}
\end{equation}
where $\mathcal{Z}$ is a set of constrained zonotopes denoted as $Z_i$, $n_z \in \mathbb{N}$ is the number of constrained zonotopes that constitute $\mathcal{Z}$, and $\mathbf{v}_i$ denotes the vertices generated from the set $\{Z_i\}$, with $n_\mathrm{verts} \in \mathbb{N}$ representing the number of vertices.

\begin{figure}
    \centering
    \includegraphics[width=0.8\linewidth]{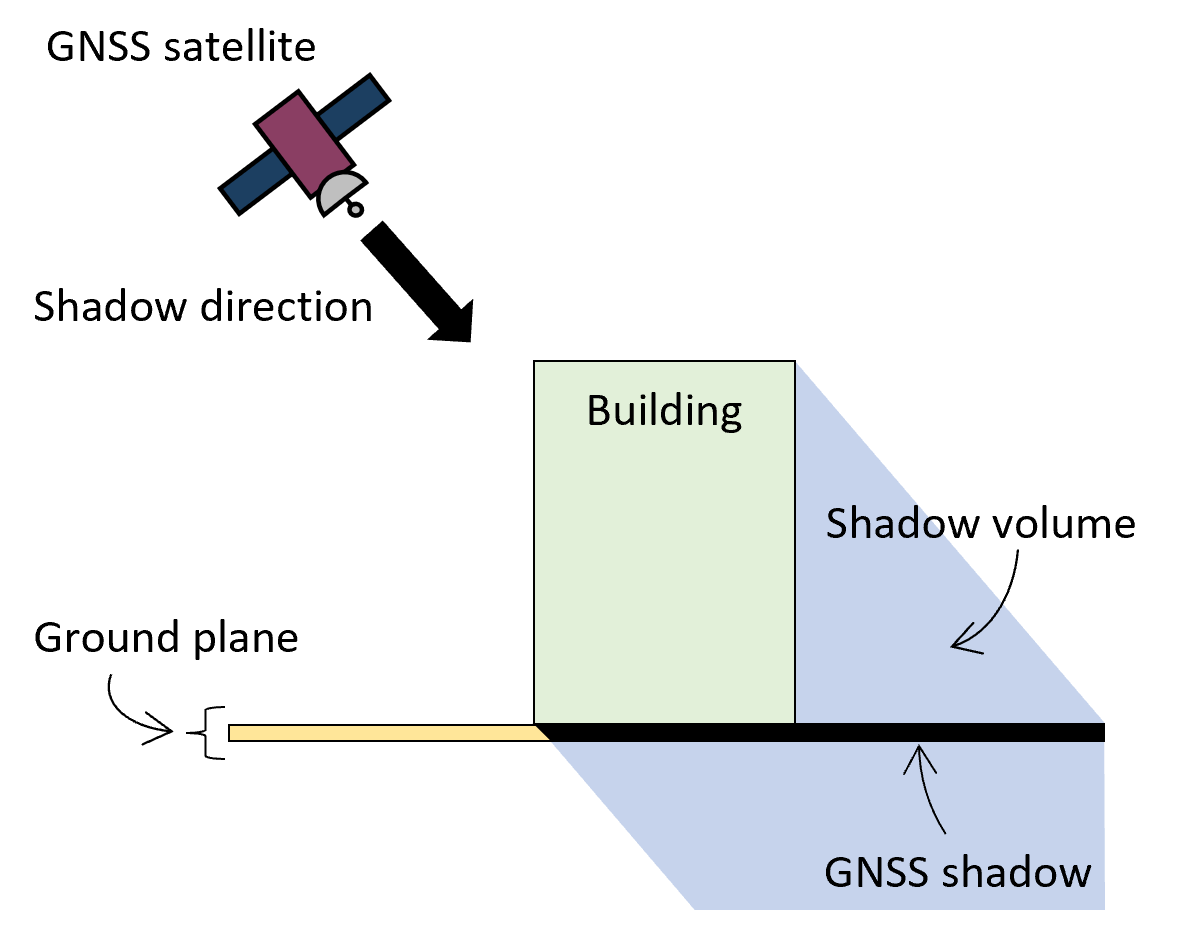}
    \caption{GNSS shadow extraction method in the existing ZSM. The shadow direction is shown with thick black arrows, the shadow volume in blue, and the GNSS shadow in black. The ground plane, shown in yellow, is fundamentally 2D but is illustrated with added thickness for visual clarity (adapted from Fig.~2 in \cite{Bhamidipati22:Set}).}
    \label{fig:GNSSShadow}
\end{figure}

\subsubsection{Computing Shadow Volumes}
\label{sec:ShadowVolumes}

The shadow volume for a satellite–building pair is generated by extending building $B_i$ in the shadow direction $\hat{\mathbf{l}}_{i,j}$. 
It is computed using the Minkowski sum of the building and the shadow direction. 
Before this, the shadow direction $\hat{\mathbf{l}}_{i,j}$ must be represented as a constrained zonotope: 
\begin{equation}
\label{eqn:ShadowDirectionZono}
L_{i,j}=\mathrm{zono}(\textbf{0}_{3\times1},\epsilon \cdot \hat{\mathbf{l}}_{i,j},[\,],[\,])
\end{equation}
where $\epsilon$ is a scaling factor for the shadow direction vector, chosen sufficiently large to ensure the construction of a complete shadow volume. 
In practice, a value of $10^5$~m was used, which far exceeds the height of the tallest building in the 3D city model.

The shadow volume $V_{i,j}$ is computed using the Minkowski sum of building $B_i$ and the shadow direction zonotope $L_{i,j}$ as follows: 
\begin{equation}
\label{eqn:ShadowVolume}
V_{i,j}^{\mathrm{shdw}} = B_i \oplus L_{i,j} = \bigcup\limits_{l=1}^{n_i}\{Z_l \oplus L_{i,j}\}
\end{equation}

\subsubsection{Computing GNSS Shadows}
\label{sec:2DShadows}

Finally, the GNSS shadow for a satellite–building pair is computed by intersecting the shadow volume $V_{i,j}^{\mathrm{shdw}}$ with the area of interest (AOI). 
The AOI represents a coarse candidate region where the receiver may be located and is defined as a subset of the ground plane: 
\begin{equation}
\label{eqn:AOI}
\begin{split}
\mathcal{A} = \{A_k\}_{k=1}^{n_\mathrm{AOI}} \subseteq \mathcal{G}
\end{split}
\end{equation}
where $A_k$ denotes a constrained zonotope within the AOI, and $n_\mathrm{AOI} \in \mathbb{N}$ is the number of constrained zonotopes representing the AOI. 
The GNSS shadow $S_{i,j}$ is then obtained by intersecting the shadow volume with the AOI as follows:
\begin{equation}
\label{eqn:2DShadow}
S_{i,j} = \bigcup\limits_{k=1}^{n_\mathrm{AOI}}\{V_{i,j}^{\mathrm{shdw}} \cap A_k\}
\end{equation}

\subsection{Estimating Set-based Receiver Position in ZSM}
\label{sec:EstimatingSet}

Here, we describe the ZSM method for estimating the set-based receiver position. 
The receiver is assumed to be located on a 2D ground plane, which simplifies the computations and is standard in conventional shadow matching \cite{Groves15:GNSS}. 
ZSM begins by defining an AOI that represents the candidate set for the receiver’s position.

The following procedure is applied to each satellite to refine the candidate set. 
\begin{itemize}
  \item First, the GNSS shadows are computed for each building based on a zonotope-represented 3D city model. 
  The GNSS shadows for each building, represented as constrained zonotopes, are then converted into vertex-based forms (i.e., polytopes). 
  This conversion is necessary to enable further set operations, such as subtraction. 
  Constrained zonotopes cannot be subtracted because they are not closed under subtraction, whereas polytopes can be subtracted since they are closed.

  \item Second, the LOS/NLOS classification of the current satellite is performed. 
  For example, Bhamidipati \textit{et al.} \cite{Bhamidipati22:Set} utilized LOS/NLOS classification based on the $C/N_0$ value. 
  If the $C/N_0$ value of the current satellite exceeds a user-specified threshold, the satellite is classified as LOS; otherwise, it is classified as NLOS. 
  More advanced classification algorithms than this simple $C/N_0$-based approach can be applied to improve ZSM performance.
  
  \item Third, the current candidate set for the receiver’s position is updated based on whether the satellite is classified as LOS or NLOS. 
  If the satellite is NLOS, the receiver is assumed to be inside the GNSS shadow, and the candidate set is updated as the intersection of the GNSS shadow with the current candidate set. 
  Conversely, if the satellite is LOS, the receiver is assumed to be outside the GNSS shadow, and the candidate set is updated by subtracting the GNSS shadow from the current candidate set.
\end{itemize}
By iteratively applying the above process to all visible satellites, the candidate set is gradually refined, ultimately yielding a 2D polytope that represents the set-based receiver position.

\section{Proposed Zonotope Shadow and Reflection Matching Algorithm}
\label{sec:Proposed}

In this section, we describe the proposed ZSRM algorithm. 
Before presenting the details, we provide an overview of the assumptions underlying the algorithm. 
We then outline the implementation methodology, including the computational procedure for GNSS reflections and the process of estimating the set-based receiver position using both GNSS shadows and reflections.

\subsection{Overview and Assumptions}
\label{sec:Overview}

The key feature of ZSRM is its ability to leverage both GNSS shadows and reflections for more accurate receiver position estimation. 
The procedure of the proposed ZSRM algorithm is as follows: 
\begin{itemize}
\item The first step is the preprocessing of the 3D city model, in which the standard 3D city model is converted into a zonotope-represented model. 
This preprocessing method is the same as that used in ZSM, as described in Section~\ref{sec:Preprocessing}.

\item The second step is the calculation of GNSS shadows and reflections. 
The method for calculating GNSS shadows is the same as that described previously for ZSM. 
In addition, ZSRM computes GNSS reflections between satellites and buildings using set operations on constrained zonotopes. 
The detailed procedure is presented in Section~\ref{sec:Reflections}.

\item The third step is the estimation of the set-based receiver position based on the satellite signal reception condition. 
Unlike ZSM, ZSRM classifies satellites into three categories: LOS-only, LOS + NLOS, and NLOS-only. 
As mentioned earlier, in ZSM, the classification between LOS and NLOS can be performed using a user-specified threshold based on the $C/N_0$ value. 
However, in ZSRM, the signal reception condition must be classified into three categories, and $C/N_0$-based classification performs poorly for this purpose. 
Several previous studies have proposed ML-based approaches that use multiple features to classify signals into LOS-only, LOS + NLOS, and NLOS-only \cite{Sun20, Kim23:Machine, Xu24}. 
In our approach, we adopted ML classifiers for signal classification, the details of which are provided in Section~\ref{sec:Performance_Realistic}. 
Based on the results of this signal classification, the receiver position is determined as follows: 
\begin{itemize}
  \item For LOS-only, the receiver is located outside the GNSS shadows and reflections. 
  \item For LOS + NLOS, the receiver is located inside the GNSS reflections. 
  \item For NLOS-only, the receiver is located inside the GNSS shadows. 
\end{itemize}
Using the above rules, ZSRM can estimate the receiver position more precisely than ZSM. 
The detailed receiver position estimation procedure is described in Section~\ref{sec:Estimating}.
\end{itemize}

Consistent with the assumptions made in the previous study \cite{Bhamidipati22:Set}, we assume that the building boundaries and road lanes in the 3D city model are accurate, and that the effects of building materials on GNSS signal propagation are neglected.

\subsection{Computing GNSS Reflections}
\label{sec:Reflections}

\begin{figure}
    \centering
    \includegraphics[width=0.85\linewidth]{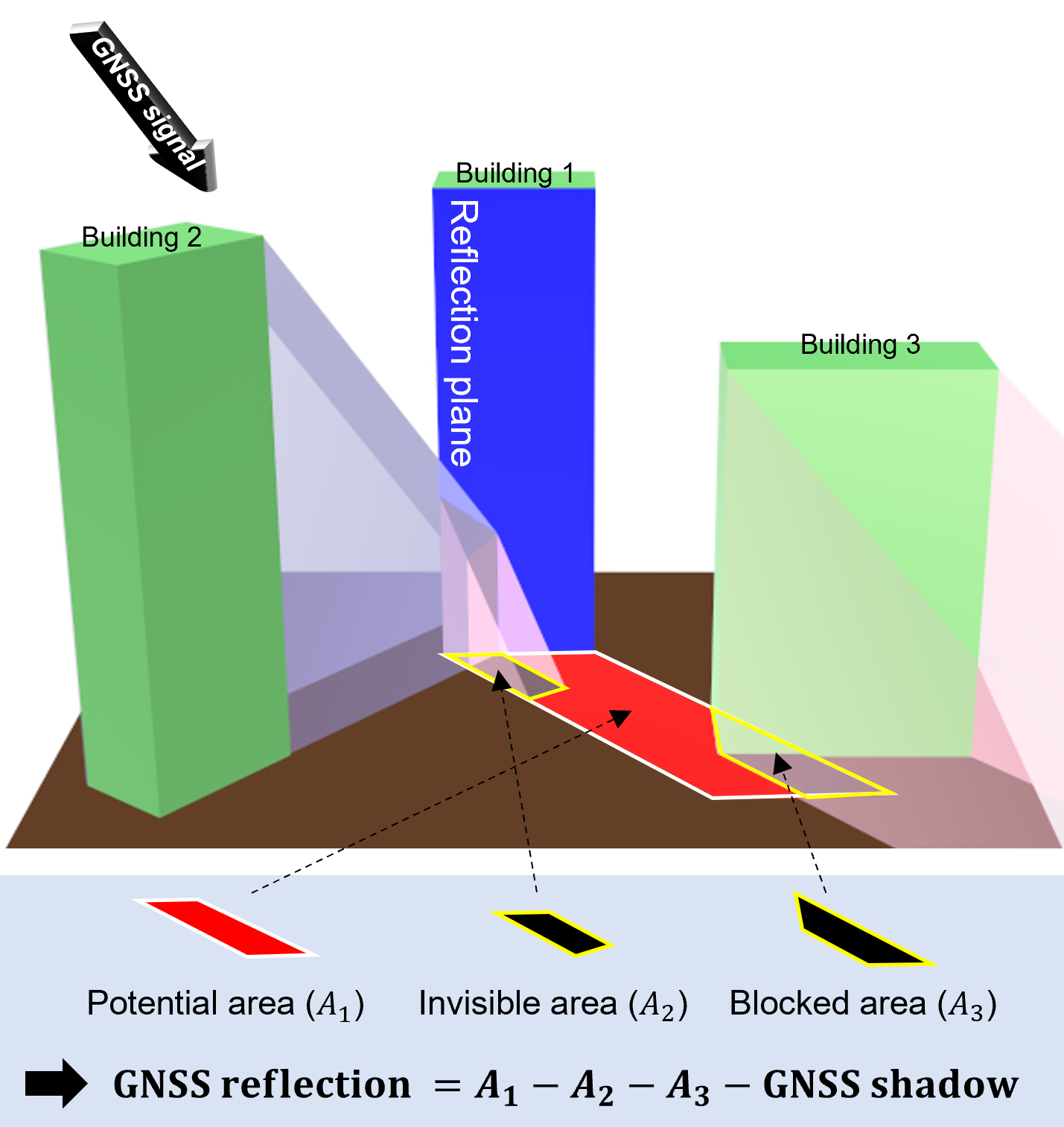}
    \caption{GNSS reflection extraction method for the proposed ZSRM. For visual clarity, only the GNSS reflections generated from a single reflection plane are shown.}
    \label{fig:CompGNSSRefl}
\end{figure}

We now describe the proposed method for computing GNSS reflections. 
A ``GNSS reflection'' is defined as an area on the ground plane where both LOS and NLOS signals from a satellite are received simultaneously. 
We assume that the receiver tracks only direct and single-reflected signals, while ignoring signals that are reflected more than once. 
This assumption is reasonable because the receiver may not be able to track multiply reflected signals owing to their weak signal strengths. 
In Fig.~\ref{fig:CompGNSSRefl}, we illustrate a simple example of the GNSS reflection computation process in 3D space. 
Since signals can be reflected from multiple planes of a building, we compute the GNSS reflection for each plane individually.

The computation method for GNSS reflections can be summarized as follows: 
\begin{itemize}
\item First, we identify the reflection planes for each satellite–building pair. 
These are the surfaces capable of reflecting signals to the ground plane. 
A detailed explanation is provided in Section~\ref{sec:ReflectionPlanes}.

\item Second, we mirror the satellite onto the reflection plane and determine the area on the ground where the reflected signals can be received. 
This corresponds to the ``potential'' area shown in Figs.~\ref{fig:CompGNSSRefl} and \ref{fig:CompCand}. 
A detailed explanation is provided in Section~\ref{sec:PotentialArea}.

\item Third, we calculate the areas where the reflected signals cannot reach due to surrounding buildings. 
These correspond to the ``invisible'' and ``blocked'' areas shown in Figs.~\ref{fig:CompGNSSRefl} and \ref{fig:CompInvBlck}. 
Signals can only be reflected by the visible surface of the reflection plane (i.e., the blue surface in Fig.~\ref{fig:CompInvBlck}(a)). 
Thus, reflections from the invisible surface of the reflection plane (i.e., the red surface in Fig.~\ref{fig:CompInvBlck}(a)) are defined as the ``invisible'' area and should be disregarded when GNSS reflections are computed. 
A detailed explanation is provided in Section~\ref{sec:InvisibleOccupied}.

\item Finally, by excluding the invisible areas, blocked areas, and GNSS shadows from the potential area, we determine the GNSS reflection, where both LOS and NLOS signals are simultaneously received. 
The GNSS shadow is also excluded to remove areas where only reflected NLOS signals are received without any LOS signal, which do not meet the definition of GNSS reflection in this study. 
Fortunately, the GNSS shadow has already been computed, allowing us to utilize the cached data without recalculation. 
A detailed explanation is provided in Section~\ref{sec:2DReflections}.
\end{itemize}

The GNSS reflection calculation process is detailed in the following subsections, in the order of: 1) finding reflection planes, 2) computing potential areas, 3) identifying invisible and blocked areas, and 4) calculating GNSS reflections.

\subsubsection{Finding Reflection Planes}
\label{sec:ReflectionPlanes}

The first step in computing GNSS reflections is to identify the set of reflection planes $\mathcal{P}_{i,j}$ for the $i$-th building and the $j$-th satellite. 
This process begins by partitioning the building into multiple planes defined as follows: 
\begin{equation}
\label{eqn:Plane}
\begin{split}
B_i = \{C_m\}_{m=1}^{n_i^\mathrm{plane}}, \, \text{where } \, C_m = \bigcup\limits_{q=1}^{n_m}Z_q
\end{split}
\end{equation}
where $C_m$ denotes a plane of building $B_i$, which consists of constrained zonotopes $Z_q$ that share the same unit normal vector $\hat{\mathbf{N}}_m$ and contain at least one vertex. 
The unit normal vector of a constrained zonotope can be obtained from its vertices. 
Furthermore, $n_i^\mathrm{plane} \in \mathbb{N}$ denotes the number of planes constituting building $B_i$, whereas $n_m \in \mathbb{N}$ denotes the number of constrained zonotopes in plane $C_m$.

\begin{figure}
    \centering
    \includegraphics[width=1.0\linewidth]{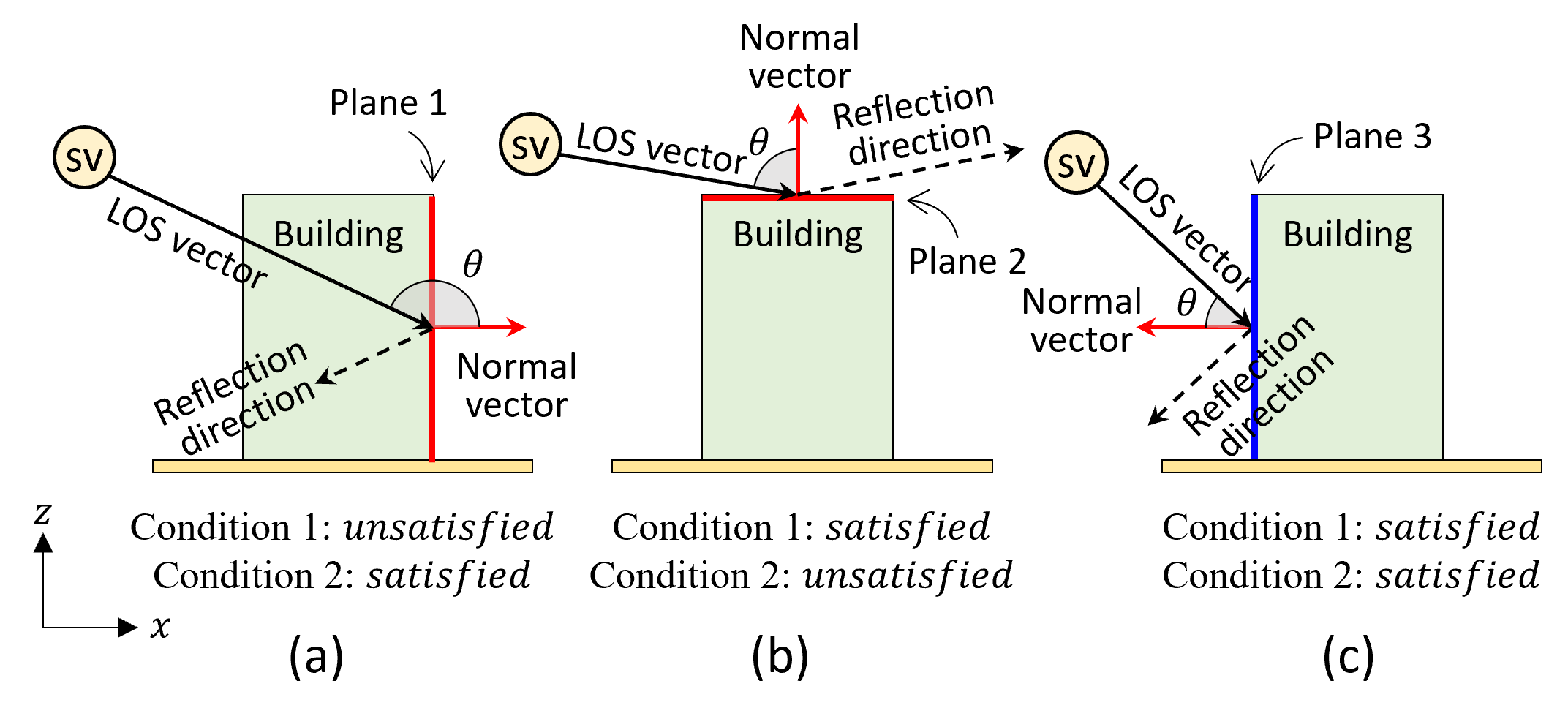}
    \caption{Example of the process of finding reflection planes for a building: (a) Plane 1 fails to satisfy Condition~1 in (\ref{eqn:FindingReflPlanes}); (b) Plane 2 fails to satisfy Condition~2 in (\ref{eqn:FindingReflPlanes}); thus, neither qualifies as a reflection plane. (c) Plane 3 satisfies both conditions, making it a valid reflection plane.}
    \label{fig:FindReflPlane}
\end{figure}

\begin{figure}
    \centering
    \includegraphics[width=0.9\linewidth]{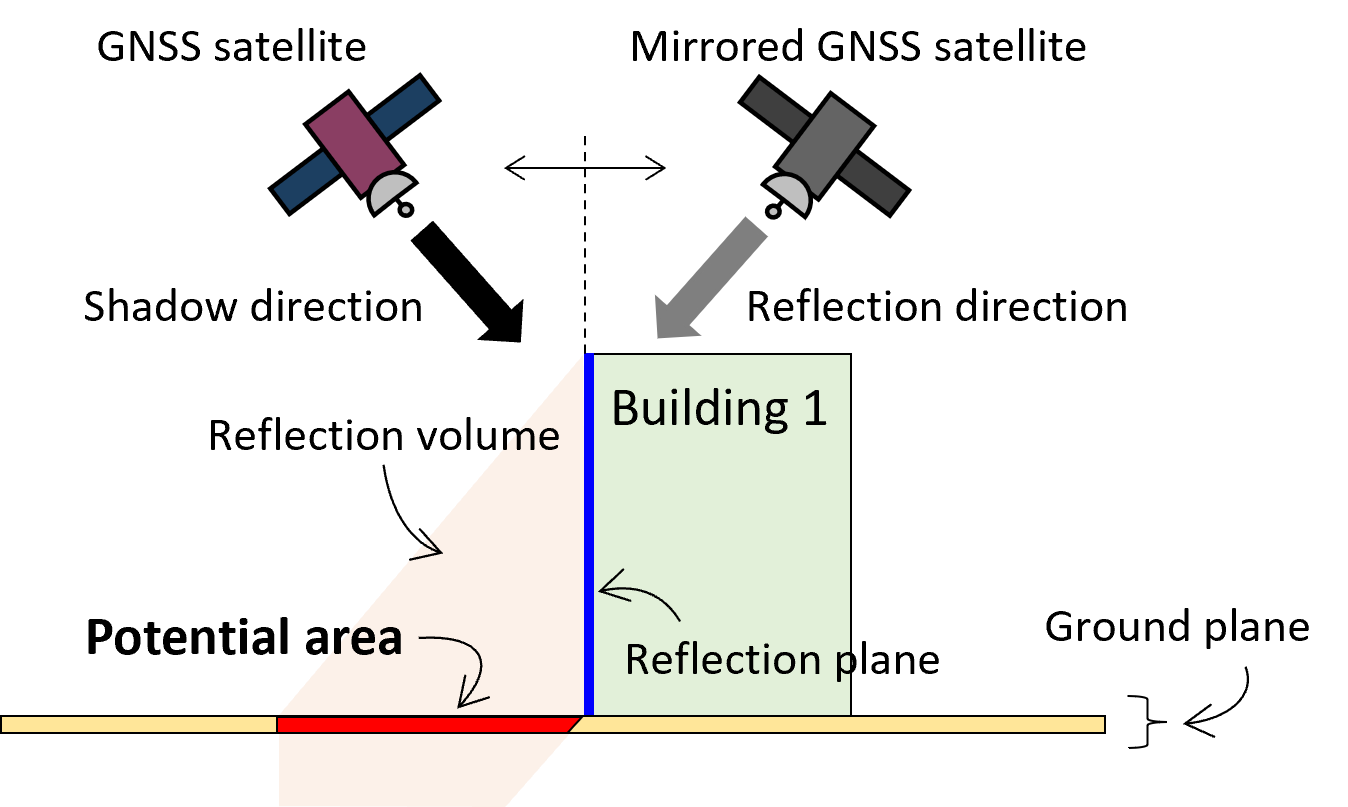}
    \caption{Computation of the ``potential'' area in ZSRM. The reflection plane is shown in blue, the reflection volume in pink, and the ``potential'' area in red. The ground plane, shown in yellow, is fundamentally 2D but is illustrated with added thickness for visual clarity.}
    \label{fig:CompCand}
\end{figure}

Subsequently, the LOS vector for the satellite–plane pair is calculated to represent the direction from the satellite to the center of the plane. 
The LOS vector $l_{m,j}^{\mathrm{los}}$ for the $m$-th plane and the $j$-th satellite is determined as follows: 
\begin{equation}
\label{eqn:LOSvector}
\hat{\mathbf{l}}_{m,j}^{\mathrm{los}}=\frac{\mathbf{c}_m-\mathbf{s}_j}{\|\mathbf{c}_m - \mathbf{s}_j\|_2}
\end{equation}
where $\mathbf{s}_j \in \mathbb{R}^3$ denotes the satellite position, and $\mathbf{c}_m \in \mathbb{R}^3$ denotes the center of plane $C_m$, calculated as the mean of the vertices of all the constrained zonotopes within $C_m$. 
The vertices of $C_m$ are obtained using (\ref{eqn:Vertices}). 
Next, we compute $\theta_{m,j}$, which represents the angle between the LOS vector $\hat{\mathbf{l}}_{m,j}^{\mathrm{los}}$ and the unit normal vector $\hat{\mathbf{N}}_m$ of the reflection plane (e.g., Fig.~\ref{fig:FindReflPlane}): 
\begin{equation}
\label{eqn:theta}
\theta_{m,j}=\mathrm{cos}^{-1}(\hat{\mathbf{N}}_m \cdot \hat{\mathbf{l}}_{m,j}^{\mathrm{los}})
\end{equation}
Furthermore, the reflection direction $\hat{\mathbf{l}}_{m,j}^{\mathrm{refl}}$, expressed as a unit vector representing the direction of the signal reflected by the plane, is calculated as follows (e.g., Fig.~\ref{fig:FindReflPlane}(c) and Fig.~\ref{fig:CompCand}): 
\begin{equation}
\label{eqn:ReflectionDirection_}
\hat{\mathbf{l}}_{m,j}^{\mathrm{refl}}=\frac{\mathbf{c}_m-\mathbf{s}_{m,j}^{\prime}}{	\|\mathbf{c}_m-\mathbf{s}_{m,j}^{\prime}\|_2}
\end{equation}
where $\mathbf{s}_{m,j}^{\prime} \in \mathbb{R}^3$ denotes the mirrored satellite position with respect to $C_m$, computed as: 
\begin{equation}
\label{eqn:MirroredSatellite}
\mathbf{s}_{m,j}^{\prime} = \mathbf{s}_{m,j} - 2(\hat{\mathbf{N}}_m \cdot (\mathbf{s}_{m,j}-\mathbf{c}_m))\hat{\mathbf{N}}_m
\end{equation}

Based on the previously computed values of $\theta_{m,j}$ and $\hat{\mathbf{l}}_{m,j}^{\mathrm{refl}}$, we determine whether plane $C_m$ qualifies as a reflection plane by verifying the following two conditions: 
\begin{equation}
\label{eqn:FindingReflPlanes}
\begin{split}
\text{Condition~1:} & \; \theta_{m,j} < 90^{\circ} \\
\text{Condition~2:} & \; \text{The $z$-component of } \hat{\mathbf{l}}_{m,j}^{\mathrm{refl}} < 0
\end{split}
\end{equation}

The first condition requires $\theta_{m,j}$ to be less than $90^\circ$, excluding the back face of the building where the satellite signal cannot reach. 
For example, in Fig.~\ref{fig:FindReflPlane}(a), when $\theta_{m,j}$ exceeds $90^\circ$, the signal cannot reach that plane (i.e., a back face of the building). 
The second condition requires that the $z$-component of the reflection direction be less than zero. 
For instance, Fig.~\ref{fig:FindReflPlane}(b) illustrates a case in which the $z$-component of the reflection direction is greater than zero. 
In this case, because the reflected signal is not directed toward the ground and therefore cannot be received by the ground receiver, the plane is excluded from consideration as a reflection plane. 
Finally, as shown in Fig.~\ref{fig:FindReflPlane}(c), the plane that satisfies both conditions is selected as the reflection plane.

We denote the set of reflection planes obtained from the $i$-th building and the $j$-th satellite as 
\begin{equation}
\label{eqn:ReflPlanes}
\begin{split}
\mathcal{P}_{i,j} = \{P_m\}_{m=1}^{n_{i,j}^\mathrm{refl}}, \, \text{where} \, P_m = \bigcup\limits_{q=1}^{n_m} Z_q \subseteq B_i
\end{split}
\end{equation}
where $P_m$ denotes a reflection plane consisting of a union of constrained zonotopes $Z_q$. 
Here, $n_{i,j}^\mathrm{refl} \in \mathbb{N}$ is the number of reflection planes for the satellite–building pair, whereas $n_m \in \mathbb{N}$ is the number of constrained zonotopes in plane $P_m$.

\subsubsection{Computing Potential Areas}
\label{sec:PotentialArea}

After finding all the reflection planes within a satellite–building pair, ``potential'' areas are calculated for each of these planes. 
In this paper, the ``potential'' area is defined as the region on the ground where reflected signals could potentially be received in the absence of surrounding buildings.
The first step in computing the potential area is to determine the mirrored position of the satellite with respect to the reflection plane. 
The mirrored satellite position is obtained using (\ref{eqn:MirroredSatellite}). 
The reflection direction, expressed as a unit vector pointing from the mirrored satellite position toward the center of the reflection plane (e.g., Fig.~\ref{fig:CompCand}), is then computed as 
\begin{equation}
\label{eqn:ReflectionDirection}
\hat{\mathbf{l}}_{m,j}^{\mathrm{refl}}=\frac{\mathbf{p}_m-\mathbf{s}_{m,j}^{\prime}}{\|\mathbf{p}_m-\mathbf{s}_{m,j}^{\prime}\|_2}
\end{equation}
where $\mathbf{s}_{m,j}^{\prime}$ denotes the mirrored satellite position, and $\mathbf{p}_m$ denotes the mean of the vertices of all constrained zonotopes constituting the reflection plane $P_m$. 
The vertices of $P_m$ are obtained using (\ref{eqn:Vertices}). 

In the same manner that the shadow direction is expressed as a constrained zonotope in (\ref{eqn:ShadowDirectionZono}), the reflection direction is represented as a constrained zonotope: 
\begin{equation}
\label{eqn:ReflectionDirectionZono}
L_{m,j}^{\mathrm{refl}}=\mathrm{zono}(\mathbf{0}_{3\times1},\epsilon \cdot \hat{\mathbf{l}}_{m,j}^{\mathrm{refl}},[\,],[\,])
\end{equation}
where $\epsilon$ is a scaling factor for the reflection direction; in practice, the same value of $10^5$ m as in (\ref{eqn:ShadowDirectionZono}) is used. 
The reflection volume $V_{m,j}^{\mathrm{refl}}$ is then computed as the Minkowski sum of the reflection plane $P_m$ and the zonotope reflection direction $L_{m,j}^{\mathrm{refl}}$: 
\begin{equation}
\label{eqn:ReflectionVolume}
V_{m,j}^{\mathrm{refl}} = P_m \oplus L_{m,j}^{\mathrm{refl}} = \bigcup\limits_{q=1}^{n_m}\{Z_q \oplus L_{m,j}^{\mathrm{refl}}\}
\end{equation}
Finally, the ``potential'' area $R_{m,j}^{\mathrm{pot}}$ is obtained by intersecting the reflection volume with the AOI defined in Section~\ref{sec:2DShadows}: 
\begin{equation}
\label{eqn:CandArea}
R_{m,j}^{\mathrm{pot}} = \bigcup\limits_{k=1}^{n_\mathrm{AOI}}\{V_{m,j}^{\mathrm{refl}} \cap A_k\}
\end{equation}

\subsubsection{Computing Invisible and Blocked Areas}
\label{sec:InvisibleOccupied}

\begin{figure}
    \centering
    \includegraphics[width=0.9\linewidth]{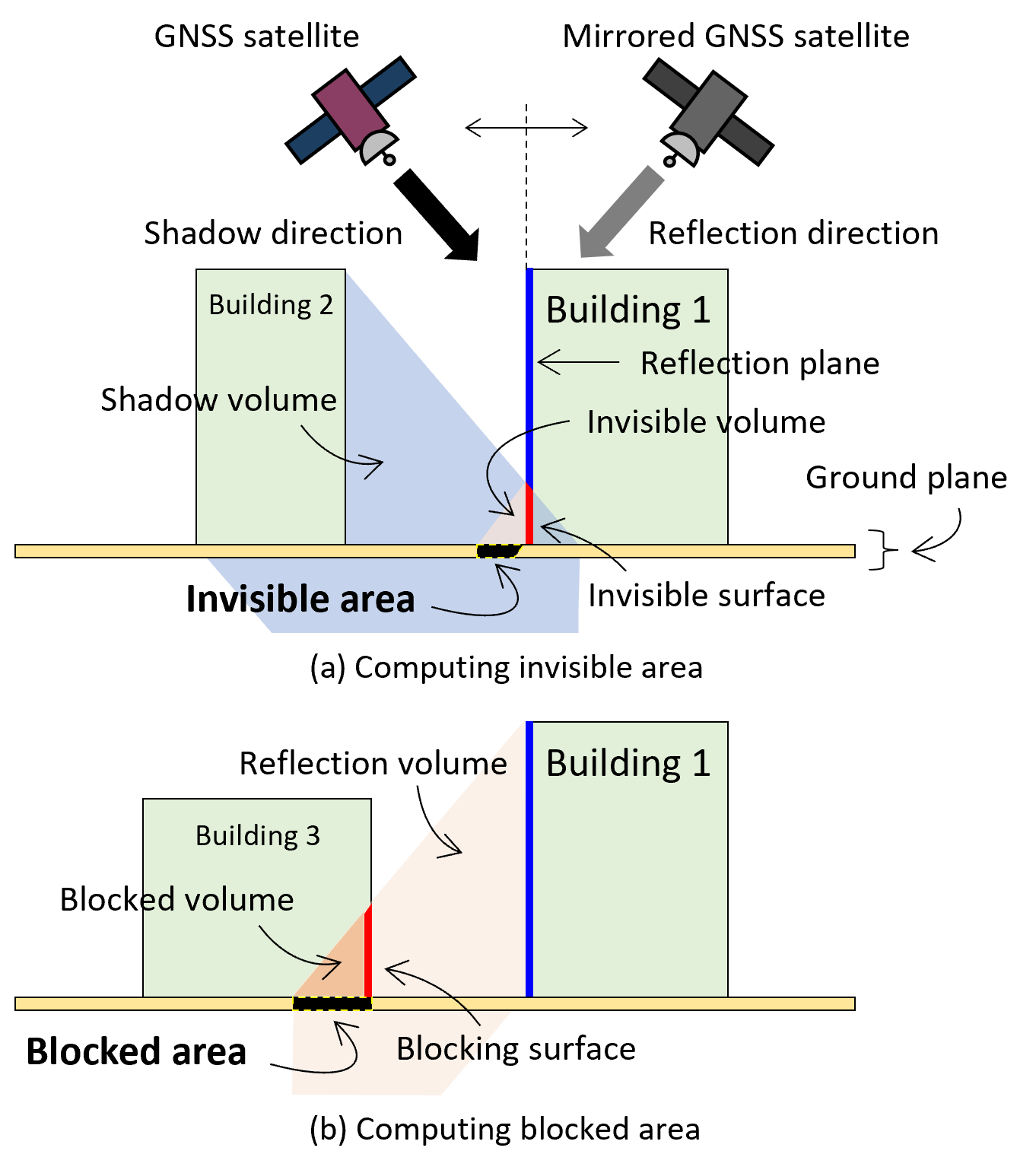}
    \caption{Computation of (a) the ``invisible'' area and (b) the ``blocked'' area in ZSRM. In (a), the shadow volume is shown in blue, the invisible surface in red, the invisible volume in pink, and the ``invisible'' area in black. In (b), the reflection volume is shown in pink, the blocking surface in red, the blocked volume in orange, and the ``blocked'' area in black. For visual clarity, the ground plane, shown in yellow, is fundamentally 2D but is illustrated with added thickness. In addition, only the invisible and blocked areas caused by a single building are depicted, although in general they are defined with respect to all buildings.}
    \label{fig:CompInvBlck}
\end{figure}

The process of computing the ``invisible'' area is illustrated in Fig.~\ref{fig:CompInvBlck}(a). 
In this paper, the ``invisible'' area is defined as the region on the ground where reflected signals from the invisible surface of the reflection plane (i.e., the red surface in Fig.~\ref{fig:CompInvBlck}(a)) would otherwise have reached.  
First, the invisible surface, which represents the segments of the reflection plane where the LOS signal is obstructed by buildings, is calculated. 
The invisible surface $S_{m,j}^{\mathrm{invis}}$ is obtained by intersecting the shadow volumes generated for all the buildings with the reflection plane $P_m$: 
\begin{equation}
\label{eqn:InvisibleSurface}
S_{m,j}^{\mathrm{invis}} = \bigcup\limits_{i=1}^{n_\mathrm{bldg}}\{V_{i,j}^{\mathrm{shdw}} \cap P_m \}
\end{equation}
where $V_{i,j}^{\mathrm{shdw}}$ is the shadow volume previously computed in (\ref{eqn:ShadowVolume}), and thus does not need to be recalculated. 
Next, the invisible volume $V_{m,j}^{\mathrm{invis}}$ is derived by computing the Minkowski sum of the invisible surface and the reflection direction: 
\begin{equation}
\label{eqn:InvisibleVolume}
V_{m,j}^{\mathrm{invis}} = S_{m,j}^{\mathrm{invis}} \oplus L_{m,j}^{\mathrm{refl}}
\end{equation}
Finally, the ``invisible'' area $R_{m,j}^{\mathrm{invis}}$ is obtained by intersecting the invisible volume with the AOI defined in Section~\ref{sec:2DShadows}: 
\begin{equation}
\label{eqn:InvisArea}
R_{m,j}^{\mathrm{invis}} = \bigcup\limits_{k=1}^{n_\mathrm{AOI}}\{ V_{m,j}^{\mathrm{invis}} \cap A_k \}
\end{equation}

Subsequently, the ``blocked'' area is computed, as shown in Fig.~\ref{fig:CompInvBlck}(b). 
In this paper, the ``blocked'' area is defined as the region on the ground where reflected signals from the reflection plane (i.e., the blue surface in Fig.~\ref{fig:CompInvBlck}(b)) would have reached if there were no blocking surface (i.e., the red surface in Fig.~\ref{fig:CompInvBlck}(b)) of a building. 
Initially, the blocking surface is obtained by intersecting the reflection volume $V_{m,j}^{\mathrm{refl}}$ with all buildings: 
\begin{equation}
\label{eqn:BlockingSurface}
S_{m,j}^{\mathrm{blck}} = V_{m,j}^{\mathrm{refl}} \cap \mathcal{B} = \bigcup\limits_{i=1}^{n_\mathrm{bldg}}\{V_{m,j}^{\mathrm{refl}} \cap B_i \}
\end{equation}
Next, the blocked volume $V_{m,j}^{\mathrm{blck}}$ is derived by computing the Minkowski sum of the blocking surface and the reflection direction: 
\begin{equation}
\label{eqn:BlockedVolume}
V_{m,j}^{\mathrm{blck}} = S_{m,j}^{\mathrm{blck}} \oplus L_{m,j}^{\mathrm{refl}}
\end{equation}
Finally, the ``blocked'' area $R_{m,j}^{\mathrm{blck}}$ is obtained by intersecting the blocked volume with the AOI defined in Section~\ref{sec:2DShadows}:  
\begin{equation}
\label{eqn:BlockedArea}
R_{m,j}^{\mathrm{blck}} = \bigcup\limits_{k=1}^{n_\mathrm{AOI}}\{V_{m,j}^{\mathrm{blck}} \cap A_k \}
\end{equation}

\subsubsection{Computing GNSS Reflections}
\label{sec:2DReflections}

To calculate the GNSS reflections (i.e., areas on the ground plane where both LOS and NLOS signals from a satellite are simultaneously received) for the $i$-th building and the $j$-th satellite, the invisible area $R_{m,j}^{\mathrm{invis}}$ and the blocked area $R_{m,j}^{\mathrm{blck}}$ are subtracted from the potential area $R_{m,j}^{\mathrm{pot}}$ for all relevant reflection planes. 
The resulting areas are then unified across all reflection planes, after which the GNSS shadow $S_{i,j}$, previously computed in (\ref{eqn:2DShadow}), is excluded.

Note that each area is represented as a set of constrained zonotopes. 
Since constrained zonotopes do not exhibit closure under subtraction, each area must be converted into a polytope of the vertex representation. 
Using (\ref{eqn:Vertices}), $R_{m,j}^{\mathrm{pot}}$, $R_{m,j}^{\mathrm{invis}}$, $R_{m,j}^{\mathrm{blck}}$, and $S_{i,j}$ are converted into vertex-representation polytopes, denoted as $R_{m,j}^{\mathrm{pot,v}}$, $R_{m,j}^{\mathrm{invis,v}}$, $R_{m,j}^{\mathrm{blck,v}}$, and $S_{i,j}^\mathrm{v}$, respectively. 
Finally, the GNSS reflection for the $i$-th satellite and the $j$-th building is computed as: 
\begin{equation}
\label{eqn:2DReflection}
R_{i,j} = \bigcup\limits_{m=1}^{n_{i,j}^\mathrm{refl}}\{R_{m,j}^{\mathrm{pot,v}} - R_{m,j}^{\mathrm{invis,v}} - R_{m,j}^{\mathrm{blck,v}}\} - S_{i,j}^\mathrm{v}
\end{equation}

\subsection{Estimating Set-Based Receiver Position in ZSRM}
\label{sec:Estimating}

\begin{figure*}
    \centering
    \includegraphics[width=0.9\linewidth]{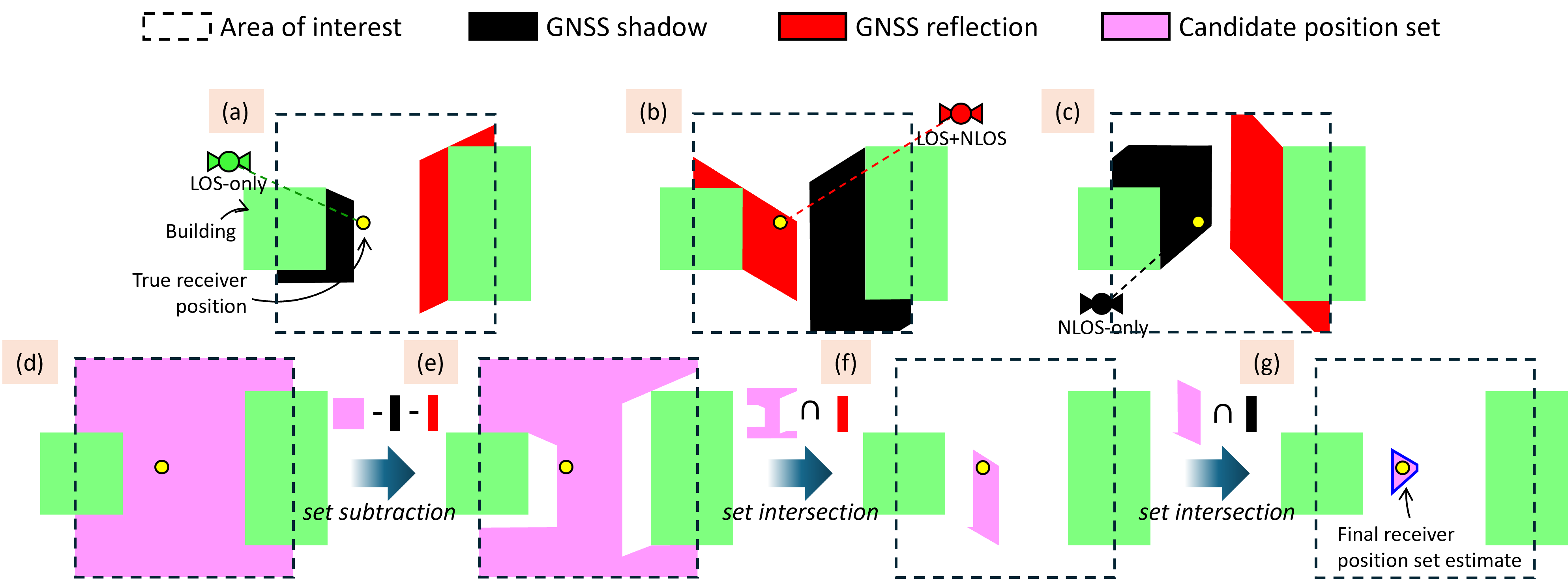}
    \caption{Simulation of estimating the receiver position set using the ZSRM algorithm. Subfigures (a)--(c) show the GNSS shadows and reflections computed for each visible satellite, while subfigures (d)--(g) illustrate the iterative refinement process of the ZSRM algorithm. The yellow circle indicates the true receiver position.}
    \label{fig:PosZSRM}
\end{figure*}

We now describe the process of estimating the set-based receiver position using the computed GNSS shadows and reflections. 
This process involves iteratively refining the coarse set-based receiver position with the GNSS shadows and reflections across all satellites. 
For clarity, a simulation of the iterative refinement process is shown in Fig.~\ref{fig:PosZSRM}, which illustrates a scenario involving two buildings and three satellites. 
In Fig.~\ref{fig:PosZSRM}, each building is shown as a green rectangle, the true receiver position as a small yellow circle, and the AOI as a transparent square outlined with a dotted line.

Figs.~\ref{fig:PosZSRM}(a)--(c) show the GNSS shadows (black) and GNSS reflections (red) computed for each satellite. 
Each satellite is depicted as an icon with a circular body and solar panels, with different colors indicating the signal reception conditions: green for LOS-only, red for LOS + NLOS, and black for NLOS-only. 
Figs.~\ref{fig:PosZSRM}(d)--(g) illustrate the iterative refinement process of the candidate receiver position set (magenta) across all satellites. 
As shown in Fig.~\ref{fig:PosZSRM}(d), the initial candidate receiver position set corresponds to the entire AOI. 
The candidate set is then refined through set operations using the GNSS shadows and reflections. 
In this process, all GNSS shadows and reflections are converted from constrained zonotopes to vertex-representation polytopes.

During the refinement process, different set operations are applied based on the signal reception conditions of the satellite, as follows: 
\begin{itemize}
  \item For LOS-only satellites, the GNSS shadows and reflections are subtracted from the current candidate position set, since a receiver receiving LOS-only signals from a satellite cannot lie within the shadows or reflections of that satellite (e.g., Fig.~\ref{fig:PosZSRM}(e)). 
  \item For LOS + NLOS satellites, the intersection between the GNSS reflections and the current candidate position set is computed, since a receiver receiving LOS + NLOS signals from a satellite must lie within the reflections, but not within the shadows, of that satellite (e.g., Fig.~\ref{fig:PosZSRM}(f)). 
  \item For NLOS-only satellites, the intersection between the GNSS shadows and the current candidate position set is computed, since a receiver receiving NLOS-only signals from a satellite must lie within the shadows, but not within the reflections, of that satellite (e.g., Fig.~\ref{fig:PosZSRM}(g)). 
\end{itemize}
After iteratively executing the above process for all satellites, the algorithm outputs a set-based receiver position on the ground plane in the form of a 2D polytope. 
The final receiver position is typically determined as the centroid of this set. 

In the example shown in Fig.~\ref{fig:PosZSRM}(g), the final receiver position set is depicted as a unified 2D polytope; however, multiple disjoint components may also be generated. 
This phenomenon, known as multi-modal ambiguity, is common in existing techniques such as conventional shadow matching and ZSM \cite{Groves15:GNSS, Bhamidipati22:Set}. 
A detailed discussion of the multi-modal ambiguity problem is presented in Section~\ref{sec:Validation}.

\subsection{ZSRM Algorithm Details}
\label{sec:ZSRM}

\newcommand\mycommfont[1]{\footnotesize\ttfamily#1}
\SetCommentSty{mycommfont}
\SetKwInput{KwInput}{Input} 
\SetKwInput{KwOutput}{Output}

\begin{algorithm*}
\label{algorithm:ZSRM}
\DontPrintSemicolon
  \KwInput{
  \begin{itemize}
  \item 3D city model including buildings $\mathcal{B} = \{B_i\}_{i=1}^{n_\mathrm{bldg}}$ and the area of interest $\mathcal{A} = \{A_k\}_{k=1}^{n_\mathrm{AOI}}$
  \item Satellite position $\mathcal{S} = \{\mathbf{s}_j\}_{j=1}^{n_\mathrm{sat}}$ and signal reception condition $\{SRC_j\}_{j=1}^{n_\mathrm{sat}}$ for each satellite
  \item Scale factor $\epsilon = 10^5$ m for shadow and reflection direction
  \end{itemize}
  }
  \KwOutput{Set-based receiver position estimate $P$}
  $P \gets \mathrm{vertices}(\mathcal{A})$ \\
  \For{each $\mathbf{s}_j \in \mathcal{S}$}{
    $C_k^\mathrm{shdw}=\emptyset$ \\
    $C_k^\mathrm{refl}=\emptyset$ \\
    \For{each $B_i \in \mathcal{B}$}{
        $\hat{\mathbf{l}}_{i,j} \gets \mathrm{makeShadowDirection}(B_i,\mathbf{s}_j)$ \\
        $L_{i,j} \gets \mathrm{zono}(\textbf{0}_{3\times1},\epsilon \cdot \hat{\mathbf{l}}_{i,j},[\,],[\,])$ \\
        \For{each $Z_l \in B_i$}{
            $V_l \gets Z_l \oplus L_{i,j}$ \\
            $V_i \gets \mathrm{append}(V_l)$ \\
            \For{each $A_k \in \mathcal{A}$}{
                $S_k \gets V_l \cap A_k$ \\
                $S_k \gets \mathrm{vertices}(\{S_k\})$ \\
                $C_k^\mathrm{shdw} \gets C_k^\mathrm{shdw} \cup S_k$ \\
            }
        }
        $V_j \gets \mathrm{append}(V_i)$ \\
    }
    \For{each $B_i \in \mathcal{B}$}{
        $\mathcal{P}_{i,j} \gets \mathrm{findReflectionPlanes}(B_i,\mathbf{s}_j)$ \\
        \For{each $P_m \in \mathcal{P}_{i,j}$}{
            $\hat{\mathbf{l}}_{m,j}^{\mathrm{refl}} \gets \mathrm{makeReflectionDirection}(P_m,\mathbf{s}_j)$ \\
            $L_{m,j}^{\mathrm{refl}} \gets \mathrm{zono}(\textbf{0}_{3\times1},\epsilon \cdot \hat{\mathbf{l}}_{m,j}^{\mathrm{refl}},[\,],[\,])$ \\
            \For{each $Z_q \in P_m$}{
                $V_q^{\mathrm{refl}} \gets Z_q \oplus L_{m,j}^{\mathrm{refl}}$ \\
                $V_q^{\mathrm{invis}} \gets \{Z_q \cap V_j\} \oplus L_{m,j}^{\mathrm{refl}}$ \\
                $V_q^{\mathrm{blck}} \gets \{V_q^{\mathrm{refl}} \cap \mathcal{B}\} \oplus L_{m,j}^{\mathrm{refl}}$ \\
                \For{each $A_k \in \mathcal{A}$}{
                    $R_k^{\mathrm{pot}} \gets V_q^{\mathrm{refl}} \cap A_k \,;$ \,\, $R_k^{\mathrm{invis}} \gets V_q^{\mathrm{invis}} \cap A_k \,;$ \,\,
                    $R_k^{\mathrm{blck}} \gets V_q^{\mathrm{blck}} \cap A_k$ \\
                    $R_k^{\mathrm{pot}} \gets \mathrm{vertices}(\{R_k^{\mathrm{pot}}\}) \,;$ \,\,
                    $R_k^{\mathrm{invis}} \gets \mathrm{vertices}(\{R_k^{\mathrm{invis}}\}) \,;$ \,\,
                    $R_k^{\mathrm{blck}} \gets \mathrm{vertices}(\{R_k^{\mathrm{blck}}\})$ \\
                    $C_k^{\mathrm{refl}} \gets C_k^{\mathrm{refl}} \cup \{R_k^{\mathrm{pot}} \setminus R_k^{\mathrm{invis}} \setminus R_k^{\mathrm{blck}} \setminus C_k^{\mathrm{shdw}}\}$ \\
                }
            }
        }
    }
    \If{$SRC_j$ is ${\mathrm{NLOS-only}}$}
    {
        $P \gets P \cap C_k^{\mathrm{shdw}}$
    }
    \ElseIf{$SRC_j$ is ${\mathrm{LOS+NLOS}}$}
    {
        $P \gets P \cap C_k^{\mathrm{refl}}$
    }
    \Else
    {
    	$P \gets P \setminus C_k^{\mathrm{shdw}} \setminus C_k^{\mathrm{refl}}$
    }
  }
\caption{ZSRM Algorithm}
\end{algorithm*}

We now provide a detailed explanation of how the ZSRM algorithm is executed in this study, as outlined in Algorithm~\ref{algorithm:ZSRM}. 
The process begins with the initial set-based receiver position estimate covering the entire AOI $\mathcal{A} \subseteq \mathcal{G}$ (line 1). 
The AOI must be defined to include the true receiver position, which is typically determined using conventional GNSS ranging-based positioning methods \cite{Groves15:GNSS}. 
In practice, we set the AOI to a 120 m $\times$ 120 m square area aligned with both the cross-street and along-street directions, centered on the elevation-weighting-based GPS single-point positioning solution. 
If no localization information is available, the AOI is set to $\mathcal{A} = \mathcal{G}$.

The following procedures are executed for each GNSS satellite (line 2): 
\begin{itemize}
  \item
  First, the GNSS shadows are computed. 
  For each building (line 5), the shadow volume is computed (lines 6--9) and cached (lines 10 and 15) for subsequent GNSS reflection calculations. 
  The shadow volume is then intersected with the AOI to derive the GNSS shadow (line 12). 
  Next, the GNSS shadow zonotope is converted into a vertex-representation polytope using (\ref{eqn:Vertices}) (line 13), followed by concatenation across all buildings (line 14). 
  
  \item
  Second, the GNSS reflections are computed. 
  For each building (line 16), the reflection planes are determined, as described in (\ref{eqn:FindingReflPlanes}) (line 17). 
  For each reflection plane (line 18), the reflection volume (line 22), invisible volume (line 23), and blocked volume (line 24) are computed. 
  Each volume is then intersected with the AOI to obtain the potential, invisible, and blocked areas (line 26). 
  Each area, expressed as a constrained zonotope, is converted into a vertex-representation polytope using (\ref{eqn:Vertices}) (line 27). 
  Finally, the remaining areas are concatenated after subtracting the invisible area, the blocked area, and the GNSS shadow from the potential area across all buildings (line 28).

  \item
  Third, set operations are performed between the current set-based position estimate and the concatenated GNSS shadows and reflections. 
  If the current satellite is NLOS-only (line 29), the concatenated GNSS shadow is intersected with the current set-based position estimate (line 30). 
  If the current satellite is LOS + NLOS (line 31), the concatenated GNSS reflection is intersected with the current set-based position estimate (line 32). 
  Otherwise, if the satellite is LOS-only (line 33), the concatenated GNSS shadows and reflections are subtracted from the current set-based position estimate (line 34).
\end{itemize}
Following the completion of the procedure for each satellite, the algorithm outputs a 2D polytope representing the set-based receiver position. 

The ZSRM algorithm is implemented in MATLAB, which employs the open-source Continuous Reachability Analyzer (CORA) toolbox \cite{Althoff15:An} for constrained zonotope representation and set operations. 
The MATLAB \texttt{polyshape} tool is used to represent the GNSS shadows and reflections as 2D polytopes and to perform set operations between them.

\section{Field Test Results} 
\label{sec:Field}

\subsection{Experimental Setup} 
\label{sec:Experimental}

\begin{figure}
    \centering
    \includegraphics[width=0.9\linewidth]{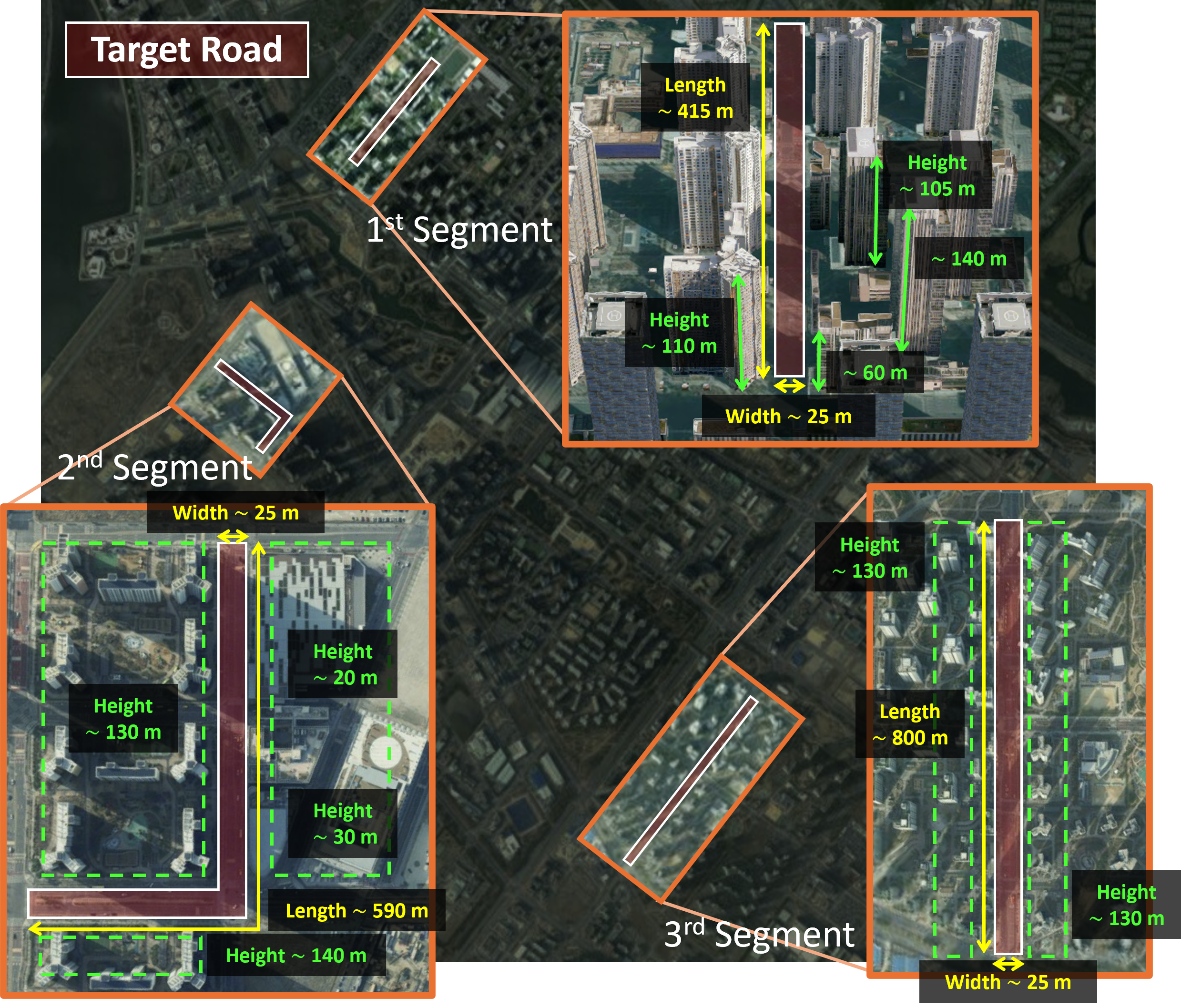}
    \caption{Target road in Songdo, surrounded by high-rise buildings.}
    \label{fig:Urban}
\end{figure}

To verify the positioning performance of the proposed ZSRM algorithm in urban areas, we conducted a field test in Songdo, Incheon, South Korea. 
As shown in Fig.~\ref{fig:Urban}, three urban road segments with distinct structural characteristics were selected to evaluate the performance of the ZSRM algorithm under diverse conditions. 
The first segment is bordered by high-rise apartment buildings and mid-rise commercial facilities on both sides. 
The second segment has high-rise residential buildings on one side and low-rise commercial buildings on the other, representing an asymmetric urban environment. 
The third segment is bordered by high-rise apartment buildings on both sides, representing a densely built-up environment. 
The total length of the road is approximately 1.8 km, with a width of 25 m.  
We used a 3D city model provided by ONEGEO \cite{ONEGEO}. 

\begin{figure}
    \centering
    \includegraphics[width=0.75\linewidth]{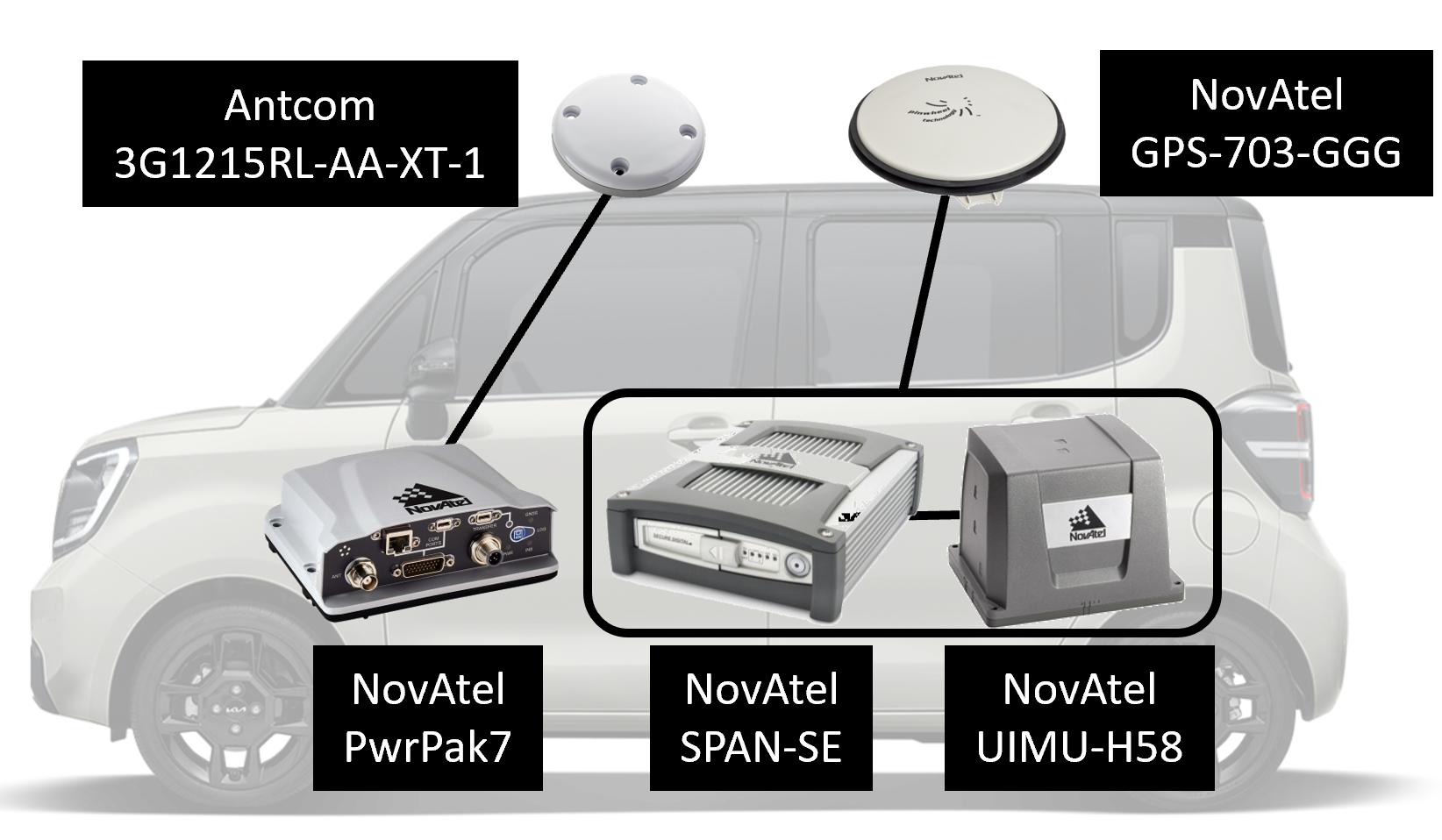}
    \caption{GNSS signal collection hardware setup.}
    \label{fig:Hardware}
\end{figure}

To collect GNSS signals, an experiment was conducted along the target road using a vehicle, as illustrated in Fig.~\ref{fig:Hardware}. 
The vehicle was equipped with an Antcom 3G1215RL-AA-XT-1 dual-polarized antenna and a NovAtel PwrPak7 GNSS receiver. 
Raw GNSS data, including both GPS and Galileo signals, were recorded on a laptop at 1 s intervals during the drive. 
The vehicle was driven under typical urban driving conditions, with speeds ranging from 0 km/h to a maximum of 37 km/h.  
To obtain the true reference trajectory, a NovAtel GPS-703-GGG antenna, NovAtel SPAN-SE, and NovAtel UIMU-H58 were installed on the vehicle. 
GNSS/INS data, acquired at 1 s intervals, were post-processed in NovAtel Inertial Explorer’s tightly coupled mode to establish the ground truth.

\begin{figure}
    \centering
    \includegraphics[width=0.75\linewidth]{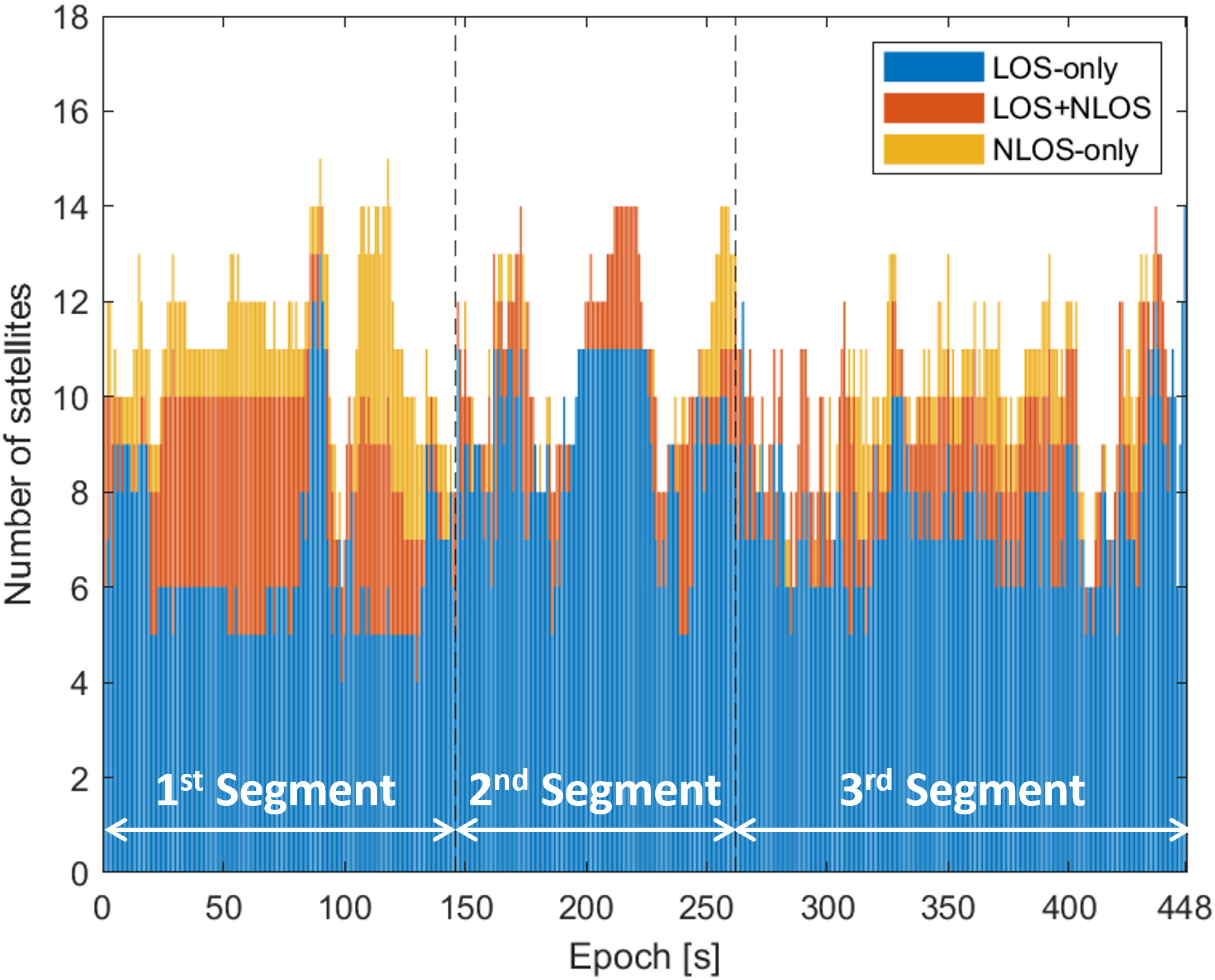}
    \caption{Number of satellites per epoch for GNSS signals collected along the target road.}
    \label{fig:SVNumber}
\end{figure}

Fig.~\ref{fig:SVNumber} presents the results of GNSS signal collection along the target road, showing the number of visible satellites per epoch, further categorized by signal reception conditions. 
Across the test route, the number of visible GNSS satellites ranged from a minimum of six to a maximum of fifteen. 
On average, the numbers of LOS-only, LOS + NLOS, and NLOS-only satellites were 7.5, 2.1, and 1.1, respectively, indicating that approximately 30\% of the signals were affected by multipath. 

\subsection{Validation Metrics and Mode Selection Algorithms} 
\label{sec:Validation}

Five validation metrics are used to evaluate performance. 
The horizontal position error is a new metric, while the remaining four are consistent with \cite{Bhamidipati22:Set}:
\begin{itemize}
  \item \textit{Horizontal position error}: the difference between the centroid of the final receiver position set and the true receiver position. 
  \item \textit{Cross-street position error}: the difference in the cross-street direction between the centroid of the final receiver position set and the true receiver position. 
  \item \textit{Along-street position error}: the difference in the along-street direction between the centroid of the final receiver position set and the true receiver position. 
  \item \textit{Cross-street position bound}: the width of the bounding box enclosing the final receiver position set in the cross-street direction. 
  \item \textit{Along-street position bound}: the width of the bounding box enclosing the final receiver position set in the along-street direction. 
\end{itemize}

\begin{figure}
    \centering
    \includegraphics[width=0.8\linewidth]{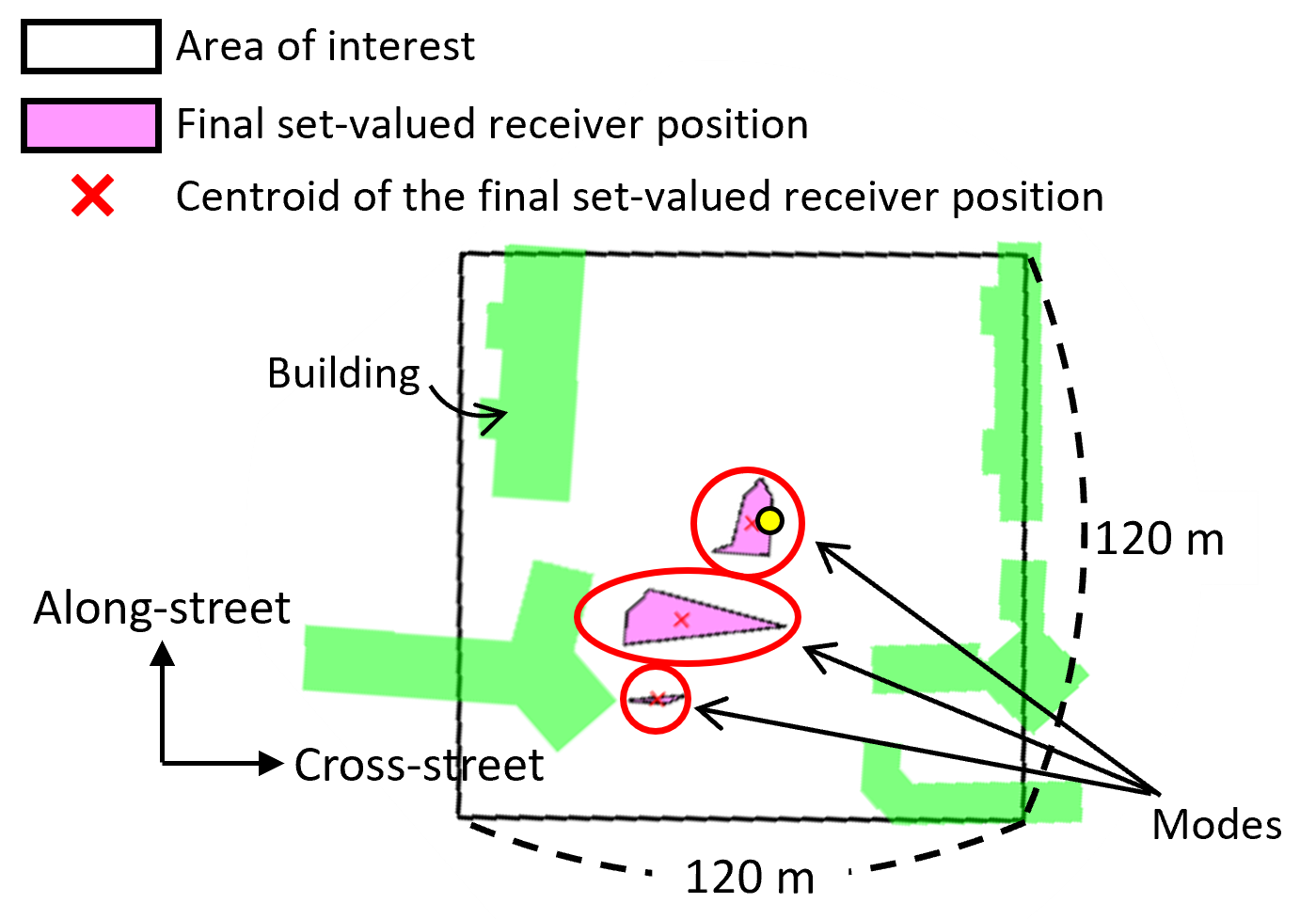}
    \caption{Illustration of the multi-modal ambiguity problem observed in both ZSM and ZSRM. This example shows a case in which the ZSRM algorithm produces multiple disjoint position sets (modes), although a similar problem can also occur in ZSM.}
    \label{fig:MultiModal}
\end{figure}

However, as illustrated in Fig.~\ref{fig:MultiModal}, both ZSM and ZSRM can generate multiple disjoint receiver position sets, each referred to as a ``mode.'' 
When multiple modes are generated, challenges arise not only in defining the receiver’s location but also in comparing the ZSM and ZSRM algorithms. 
To mitigate such multi-modal ambiguity, Neamati \emph{et al.} \cite{Neamati22:Set} proposed a mode selection algorithm called the Satellite-Pseudorange Consistency (SPC) filter. 
In the SPC filter, the mode distribution in the 2D position domain is transformed into the range-offset domain using the SPC plane constructed from the pseudoranges of each satellite. 
A mixture model is then created by fusing the range-offset data from all satellites, with higher weights assigned to LOS satellites (i.e., LOS-only or LOS + NLOS).

This approach inherently relies on an ML-based LOS classifier to provide probabilistic estimates that serve as the mixture weights. 
In our study, since the ZSRM algorithm already employs an ML classifier to categorize each satellite signal into LOS-only, LOS + NLOS, or NLOS-only, we directly reused the classifier’s output. 
Specifically, the LOS probability used in the SPC filter was computed by summing the predicted probabilities of the LOS-only and LOS + NLOS classes.  
Subsequently, $k$ range-offset samples were randomly drawn from the mixture model, and the Dirichlet distribution was computed for each mode to select the mode with the highest likelihood. 
In practice, $k$ was set to 5000. 
In our experiments, the SPC filter–based mode selection algorithm was applied to both ZSM and ZSRM to resolve multi-modal ambiguity and determine the final receiver position set. 

\subsection{Performance Comparison under Ideal Classification Conditions} 
\label{sec:Performance_Ideal} 

Both ZSM and ZSRM refine the receiver’s position by applying geometric constraints---such as GNSS shadows and reflections---based on classified satellite signals to the initial coarse candidate set. 
Misclassified signals can introduce erroneous constraints, potentially excluding the true receiver position and causing estimation failure. 
While the previous study \cite{Bhamidipati22:Set} evaluated ZSM under ideal conditions assuming perfect classification, this study analyzes the performance of both ZSM and ZSRM under ideal conditions as well as under realistic conditions that reflect classification errors observed in practice.

In this subsection, we analyze the performance of ZSM and ZSRM under the assumption of an ideal classifier. 
This analysis isolates the impact of the positioning algorithm itself, independent of classification errors. 
For this purpose, the true signal reception conditions (i.e., LOS-only, LOS + NLOS, or NLOS-only) are assigned using the true receiver position, which is available only in this controlled evaluation. 
We emphasize that this setup does not represent a practical scenario, but is designed to establish a theoretical upper bound on the achievable performance of each algorithm under perfect signal classification. 
This evaluation clarifies the extent to which performance improvement can be attributed purely to the design of the positioning algorithm, independent of classification accuracy.

To compare the positioning performance of ZSM and ZSRM, we considered two scenarios. 
The first scenario assumes ideal mode selection, in which the mode containing the true receiver position is always selected. 
Although the assumption of ideal mode selection does not reflect real-world constraints, this evaluation provides a theoretical benchmark for understanding the maximum potential improvement achievable by ZSRM over ZSM when the correct mode is chosen. 
As summarized in Table~\ref{tab:Ideal1}, ZSM yielded an RMS horizontal position error of 17.5 m, whereas ZSRM reduced this to 7.1 m, indicating a 59.7\% improvement. 
Regarding RMS position bounds, ZSM recorded 51.2 m and 67.5 m in the cross-street and along-street directions, respectively, while ZSRM improved these to 25.5 m and 27.5 m, corresponding to improvements of 50.1\% and 59.3\%.

\begin{table*}
\centering
\caption{RMS position errors and bounds of ZSM and ZSRM under an ideal classifier with ideal mode selection}
\label{tab:Ideal1}
\vspace{-2mm}
\renewcommand{\arraystretch}{1.0}
\begin{tabular}{>{\centering\arraybackslash}m{3.6cm}
                >{\centering\arraybackslash}m{2.0cm}
                >{\centering\arraybackslash}m{2.0cm}
                >{\centering\arraybackslash}m{2.0cm}
                >{\centering\arraybackslash}m{2.0cm}
                >{\centering\arraybackslash}m{2.0cm}}
\toprule
{} & 
\begin{tabular}{@{}c@{}}\normalfont Horizontal\\\normalfont position error\end{tabular} &
\begin{tabular}{@{}c@{}}\normalfont Cross-street\\\normalfont position error\end{tabular} &
\begin{tabular}{@{}c@{}}\normalfont Along-street\\\normalfont position error\end{tabular} &
\begin{tabular}{@{}c@{}}\normalfont Cross-street\\\normalfont position bound\end{tabular} &
\begin{tabular}{@{}c@{}}\normalfont Along-street\\\normalfont position bound\end{tabular} \\
\midrule
\end{tabular}
\renewcommand{\arraystretch}{1.7}
\begin{tabular}{>{\centering\arraybackslash}m{3.6cm}
                >{\centering\arraybackslash}m{2.0cm}
                >{\centering\arraybackslash}m{2.0cm}
                >{\centering\arraybackslash}m{2.0cm}
                >{\centering\arraybackslash}m{2.0cm}
                >{\centering\arraybackslash}m{2.0cm}}
ZSM (Existing method \cite{Bhamidipati22:Set})   & 17.5 m & 11.1 m & 13.6 m & 51.2 m & 67.5 m \\
\rowcolor{gray!20}
ZSRM (Proposed method)  & 7.1 m & 4.6 m & 5.3 m & 25.5 m & 27.5 m \\
Improvement             & 59.7\% & 58.3\% & 60.6\% & 50.1\% & 59.3\% \\
\bottomrule
\end{tabular}
\end{table*}

In contrast, the second scenario reflects a more practical setting by employing the SPC filter–based mode selection algorithm to choose the most likely mode in a multi-modal ambiguity environment. 
Unlike the ideal case, this approach allows for the possibility of selecting an incorrect mode, which may not contain the true receiver position and thus increases the risk of larger errors. 
As shown in Table~\ref{tab:Ideal2}, both algorithms exhibited higher errors than in the ideal case. 
Nevertheless, ZSRM still outperformed ZSM, achieving a 24.3\% reduction in RMS horizontal position error and 47.2\%--56.8\% improvements in RMS position bounds.

\begin{table*}
\centering
\caption{RMS position errors and bounds of ZSM and ZSRM under an ideal classifier with SPC filter–based mode selection}
\label{tab:Ideal2}
\vspace{-2mm}
\renewcommand{\arraystretch}{1.0}
\begin{tabular}{>{\centering\arraybackslash}m{3.6cm}
                >{\centering\arraybackslash}m{2.0cm}
                >{\centering\arraybackslash}m{2.0cm}
                >{\centering\arraybackslash}m{2.0cm}
                >{\centering\arraybackslash}m{2.0cm}
                >{\centering\arraybackslash}m{2.0cm}}
\toprule
{} & 
\begin{tabular}{@{}c@{}}\normalfont Horizontal\\\normalfont position error\end{tabular} &
\begin{tabular}{@{}c@{}}\normalfont Cross-street\\\normalfont position error\end{tabular} &
\begin{tabular}{@{}c@{}}\normalfont Along-street\\\normalfont position error\end{tabular} &
\begin{tabular}{@{}c@{}}\normalfont Cross-street\\\normalfont position bound\end{tabular} &
\begin{tabular}{@{}c@{}}\normalfont Along-street\\\normalfont position bound\end{tabular} \\
\midrule
\end{tabular}
\renewcommand{\arraystretch}{1.7}
\begin{tabular}{>{\centering\arraybackslash}m{3.6cm}
                >{\centering\arraybackslash}m{2.0cm}
                >{\centering\arraybackslash}m{2.0cm}
                >{\centering\arraybackslash}m{2.0cm}
                >{\centering\arraybackslash}m{2.0cm}
                >{\centering\arraybackslash}m{2.0cm}}
ZSM (Existing method \cite{Bhamidipati22:Set})   & 20.4 m & 14.4 m & 14.5 m & 51.4 m & 69.0 m \\
\rowcolor{gray!20}
ZSRM (Proposed method)  & 15.4 m & 11.4 m & 10.4 m & 27.2 m & 29.8 m \\
Improvement             & 24.3\% & 20.7\% & 28.2\% & 47.2\% & 56.8\% \\
\bottomrule
\end{tabular}
\end{table*}

This reduced gain can be attributed to the lower accuracy of the SPC filter in selecting the correct mode when applied to ZSRM. 
The decreased mode selection accuracy arises from two key characteristics of ZSRM’s mode distribution. 
First, ZSRM tends to generate more modes than ZSM, owing to its use of both GNSS shadow and reflection information for position refinement. 
On average, ZSM produced 2.12 modes per epoch, whereas ZSRM produced 3.09. 
Second, ZSRM often produces several incorrect modes located in close proximity to the correct mode. 
The average distance between the correct mode and the nearest incorrect mode was 60.3 m for ZSM and 24.4 m for ZSRM, indicating that ZSRM generates many small, clustered modes. 
This clustering increases ambiguity in mode selection, making it more difficult for the SPC filter to distinguish the correct mode from nearby incorrect ones. 
Consequently, the correct mode was selected in 85\% of the epochs for ZSM but only in 75\% of the epochs for ZSRM. 

\begin{figure}
    \centering
    \includegraphics[width=0.95\linewidth]{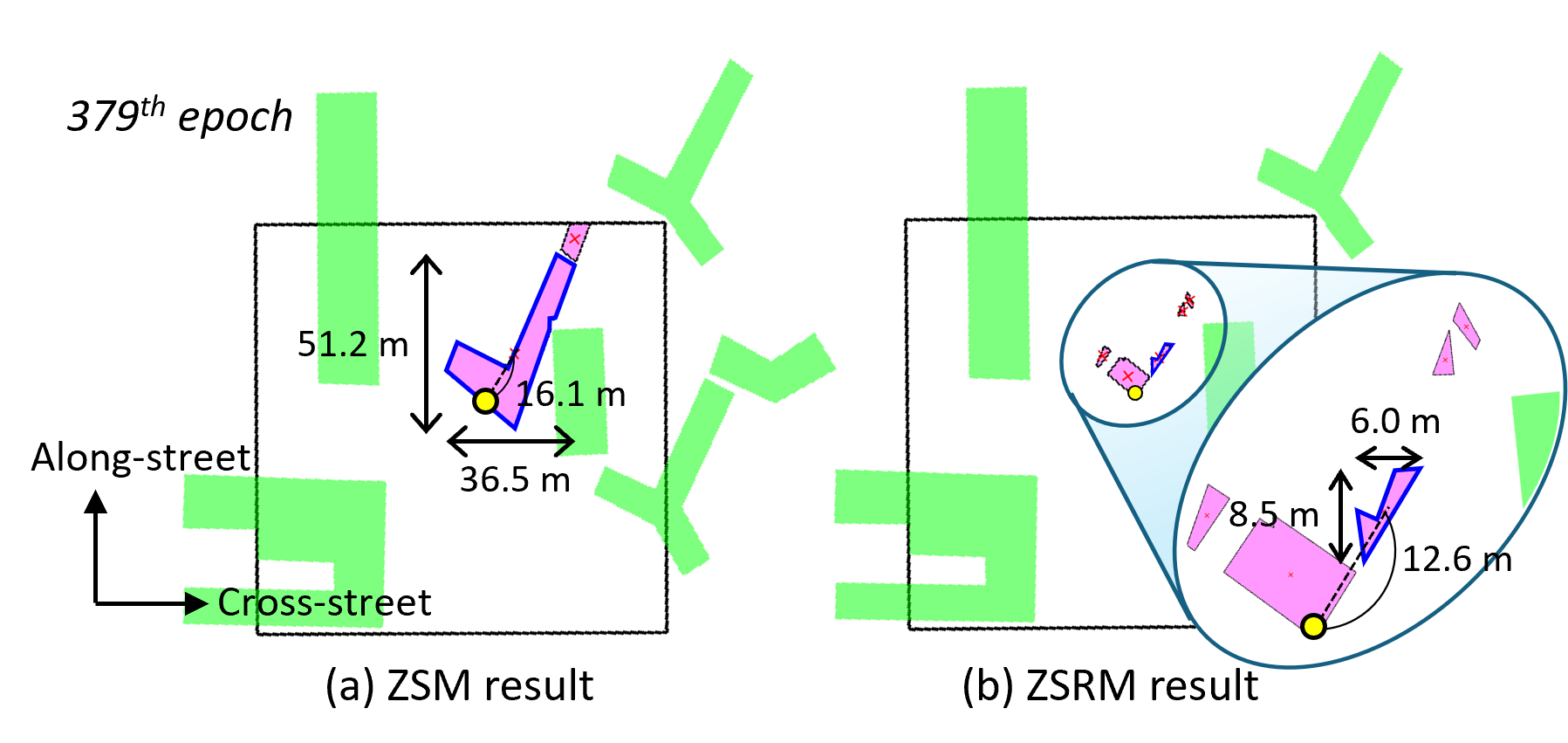}
    \caption{Results of ZSM and ZSRM at the 379\textsuperscript{th} epoch. Modes outlined in blue represent those selected by the SPC filter.}
    \label{fig:Ideal_result2}
\end{figure}

Fig.~\ref{fig:Ideal_result2} illustrates an example in which the SPC filter selected an incorrect mode when applied to ZSRM. 
The figure shows the results of ZSM and ZSRM at the 379\textsuperscript{th} epoch, where ZSM produced two modes and ZSRM produced five. 
The modes outlined in blue indicate those selected by the SPC filter. 
As shown in Fig.~\ref{fig:Ideal_result2}(a), the SPC filter correctly selected the mode containing the receiver in the ZSM case, whereas in Fig.~\ref{fig:Ideal_result2}(b), it selected an incorrect mode that did not include the receiver in the ZSRM case. 
Notably, although an incorrect mode was selected in this ZSRM case, it was the one closest to the correct mode---a behavior consistently observed across many epochs.

\begin{figure}
    \centering
    \includegraphics[width=0.6\linewidth]{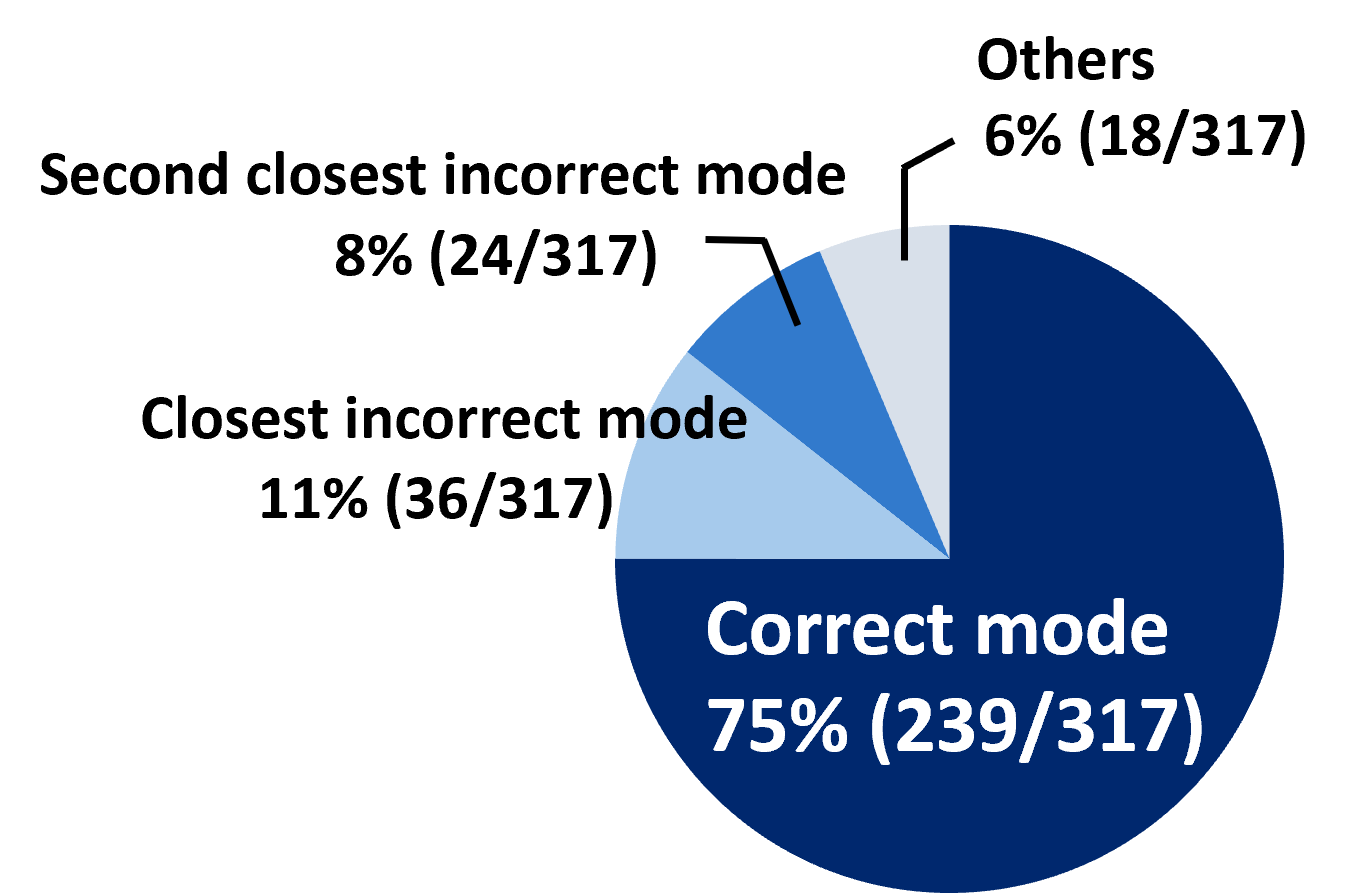}
    \caption{Distribution of mode selection results in ZSRM using the SPC filter.}
    \label{fig:SPC_analysis}
\end{figure}

Fig.~\ref{fig:SPC_analysis} summarizes the distribution of mode selection results in ZSRM. 
The correct mode was selected in 75\% of the epochs, while the closest and second-closest incorrect modes were selected in 11\% and 8\% of the epochs, respectively. 
Because of this tendency, even when incorrect modes were selected, the resulting position error did not increase significantly, enabling ZSRM to maintain lower overall position errors than ZSM. 
These results demonstrate the robustness of ZSRM in practical multi-modal environments, where perfect mode selection cannot always be guaranteed.

\subsection{Performance Comparison under Realistic Classification Conditions} 
\label{sec:Performance_Realistic} 

\begin{table*}
\centering
\caption{Confusion matrices for different classification methods---RF, GBDT, SVM, and their unanimous voting---showing the number of signals classified per class.}
\renewcommand{\arraystretch}{1.2}
\newcolumntype{C}[1]{>{\centering\arraybackslash}b{#1}}
\begin{tabular}{|C{2.2cm}|C{0.15cm}|*{3}{C{0.6cm}}|*{3}{C{0.6cm}}|*{3}{C{0.6cm}}|*{3}{C{0.6cm}}|}
\hline
Algorithms & 
& \multicolumn{3}{c|}{RF} 
& \multicolumn{3}{c|}{GBDT} 
& \multicolumn{3}{c|}{SVM} 
& \multicolumn{3}{c|}{\begin{tabular}[c]{@{}c@{}}Unanimous voting\\(RF + GBDT + SVM)\end{tabular}} \\
\cline{1-14}
Predicted class & & 0 & 1 & 2 & 0 & 1 & 2 & 0 & 1 & 2 & 0 & 1 & 2 \\
\hline
\multirow{3}{*}{Actual class} & 0 & 241 & 154 & 74 & 318 & 110 & 41 & 321 & 58 & 90 & 191 & 25 & 17 \\
& 1 & 333 & 2588 & 453 & 323 & 2569 & 482 & 518 & 2070 & 786 & 128 & 1838 & 194 \\
& 2 & 185 & 559 & 210 & 341 & 326 & 287 & 261 & 183 & 510 & 61 & 138 & 104 \\
\hline
Accuracy &
& \multicolumn{3}{c|}{63.4\%}
& \multicolumn{3}{c|}{66.2\%}
& \multicolumn{3}{c|}{60.5\%}
& \multicolumn{3}{c|}{78.8\%} \\
\hline
Class accuracy & 
& 51.4\% & 76.7\% & 22.0\% 
& 67.8\% & 76.1\% & 30.1\% 
& 68.4\% & 61.4\% & 53.5\% 
& 78.3\% & 85.1\% & 34.3\% \\
\hline
\end{tabular}
\vspace{1mm} 
\begin{flushleft} 
\hspace{12mm} \scriptsize{* Classes 0, 1, and 2 denote NLOS-only, LOS-only, and LOS + NLOS, respectively.} 
\end{flushleft}
\label{tab:ClassifierResult}
\end{table*}

This subsection compares the positioning performance of ZSM and ZSRM under realistic conditions. 
In this case, GNSS signals are classified using trained ML classifiers, and the performance of both algorithms is evaluated. 
To ensure a fair comparison, the same classifier is applied to both ZSM and ZSRM. 
Specifically, the classifier is trained to perform three-class classification (LOS-only, LOS + NLOS, and NLOS-only). 
For application to ZSM, classifier outputs labeled as either LOS-only or LOS + NLOS are both treated as LOS, since ZSM requires only a binary classification between LOS and NLOS signals, consistent with the previous study \cite{Bhamidipati22:Set}.

As mentioned earlier, misclassified satellites can introduce incorrect geometric constraints, potentially leading to positioning failures. 
In our field tests, when three different ML classifiers---random forest (RF), gradient boosting decision tree (GBDT), and support vector machine (SVM)---were individually applied, ZSM failed to produce a receiver position set in 28\%--47\% of epochs, whereas ZSRM failed in 70\%--75\% of epochs. 
The higher failure rate of ZSRM can be attributed to the larger number of misclassified satellites resulting from its more detailed three-class classification scheme, as opposed to the two-class classification used in ZSM. 
These frequent positioning failures hinder a meaningful comparison between ZSM and ZSRM.

To address this issue and improve the robustness of both algorithms, we adopted a satellite selection strategy based on classifier consensus. 
Specifically, we applied a unanimous voting strategy, in which only satellites classified identically by all three classifiers were used for positioning. 
This strategy, proposed in our previous work \cite{Kim25:Enhancing}, was shown to enhance the robustness of ZSM, and in this study, it was extended to both ZSM and ZSRM. 
The results demonstrate that unanimous voting significantly reduced the positioning failure rates to 3\% for ZSM and 4\% for ZSRM. 
Accordingly, the unanimous voting strategy was applied to both algorithms, and the performance comparison was conducted using only the satellites selected through this strategy.

\begin{figure}
    \centering
    \includegraphics[width=0.9\linewidth]{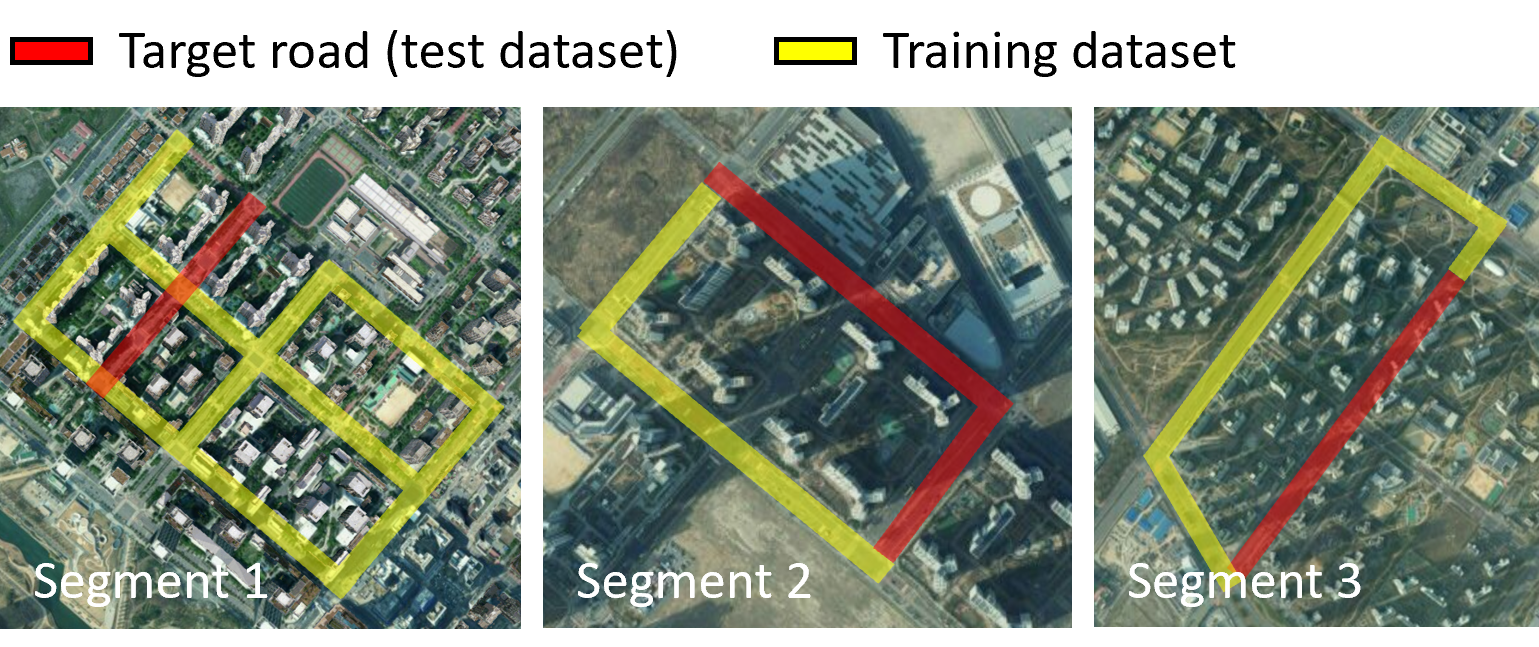}
    \caption{Training dataset collection areas.}
    \label{fig:Training_dataset}
\end{figure}

We now describe the three ML classifiers used in this study. 
The classifiers were trained using GNSS signals collected in an environment different from the target road, as shown in Fig.~\ref{fig:Training_dataset}. 
The input features for classification were $C/N_0$, elevation angle, and pseudorange residual. 
Classification results on the target road signals are summarized in Table~\ref{tab:ClassifierResult}, which presents the confusion matrices for the three-class setting. 
The overall classification accuracy ranged from 60.5\% to 66.2\%, with the poorest performance observed for LOS + NLOS signals. 
When unanimous voting was applied, 56.4\% of the signals were retained, and the classification accuracy improved to 78.8\%. 
The number of misclassified satellites per epoch decreased from 3.62--4.23 with individual classifiers to 1.26 with unanimous voting. 
In the binary classification setting, the overall accuracy ranged from 80.7\% to 84.4\%, and unanimous voting further improved the accuracy to 92.3\%, while retaining 74.8\% of the signals.

The performance comparison results of ZSM and ZSRM are presented in Table~\ref{tab:Realistic}. 
These results correspond to the scenario in which positioning is performed using only the satellites selected through unanimous voting, with mode selection in multi-modal ambiguity situations conducted using the SPC filter. 
The RMS horizontal position errors for ZSM and ZSRM were 35.9 m and 32.3 m, respectively, both higher than those observed under ideal classification. 
This degradation is primarily due to classification errors; however, the slightly larger performance drop in ZSRM can be attributed to the increased difficulty of three-class classification compared with the binary classification used in ZSM.

\begin{table*}
\centering
\caption{Comparison of RMS position errors and bounds for ZSM and ZSRM with ML classifiers using a unanimous voting strategy under the SPC filter–based mode selection scenario}
\label{tab:Realistic}
\vspace{-2mm}
\renewcommand{\arraystretch}{1.0}
\begin{tabular}{>{\centering\arraybackslash}m{3.6cm}
                >{\centering\arraybackslash}m{2.0cm}
                >{\centering\arraybackslash}m{2.0cm}
                >{\centering\arraybackslash}m{2.0cm}
                >{\centering\arraybackslash}m{2.0cm}
                >{\centering\arraybackslash}m{2.0cm}}
\toprule
{} & 
\begin{tabular}{@{}c@{}}\normalfont Horizontal\\\normalfont position error\end{tabular} &
\begin{tabular}{@{}c@{}}\normalfont Cross-street\\\normalfont position error\end{tabular} &
\begin{tabular}{@{}c@{}}\normalfont Along-street\\\normalfont position error\end{tabular} &
\begin{tabular}{@{}c@{}}\normalfont Cross-street\\\normalfont position bound\end{tabular} &
\begin{tabular}{@{}c@{}}\normalfont Along-street\\\normalfont position bound\end{tabular} \\
\midrule
\end{tabular}
\renewcommand{\arraystretch}{1.7}
\begin{tabular}{>{\centering\arraybackslash}m{3.6cm}
                >{\centering\arraybackslash}m{2.0cm}
                >{\centering\arraybackslash}m{2.0cm}
                >{\centering\arraybackslash}m{2.0cm}
                >{\centering\arraybackslash}m{2.0cm}
                >{\centering\arraybackslash}m{2.0cm}}
ZSM (Existing method \cite{Bhamidipati22:Set})   & 35.9 m & 25.9 m & 24.9 m & 60.6 m & 78.0 m \\
\rowcolor{gray!20}
ZSRM (Proposed method)  & 32.3 m & 24.6 m & 21.0 m & 49.7 m & 54.1 m \\
Improvement             & 10.0\% & 5.0\% & 15.7\% & 18.0\% & 30.7\% \\
\bottomrule
\end{tabular}
\end{table*}

\begin{figure}
    \centering
    \includegraphics[width=0.95\linewidth]{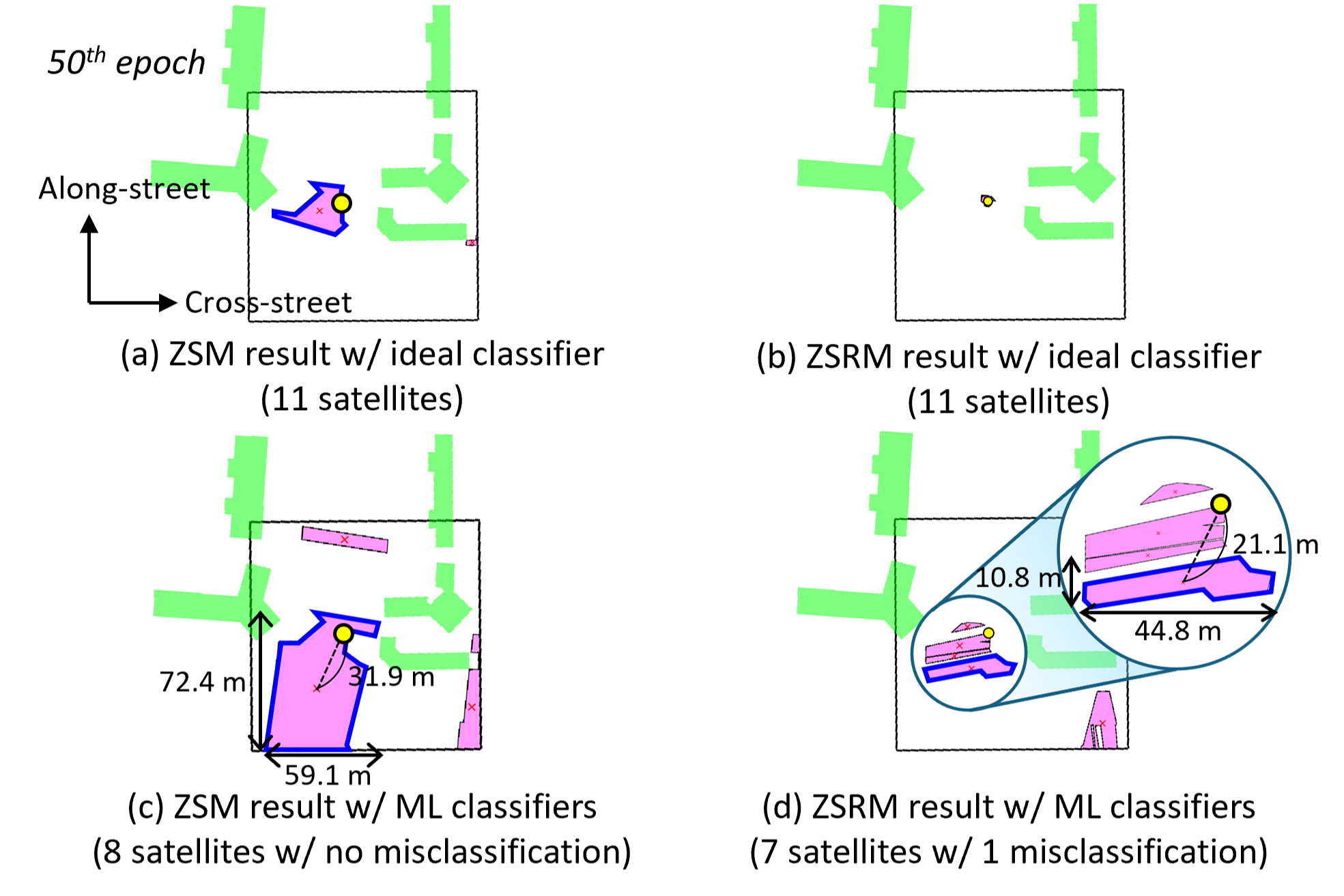}
    \caption{Results of ZSM and ZSRM at the 50\textsuperscript{th} epoch. Modes outlined in blue represent those selected by the SPC filter.}
    \label{fig:ML_result}
\end{figure}

Fig.~\ref{fig:ML_result} illustrates the mode distributions of ZSM and ZSRM for the 50\textsuperscript{th} epoch. 
Figs.~\ref{fig:ML_result}(a) and \ref{fig:ML_result}(b) show the results under ideal classification, where the signal reception conditions of all satellites are assumed to be perfectly known. 
In this case, 11 satellites were used for both ZSM and ZSRM. 
Figs.~\ref{fig:ML_result}(c) and \ref{fig:ML_result}(d) show the results using only the satellites selected through unanimous voting based on ML classifiers. 
For ZSM, 8 satellites were selected, all correctly classified. 
For ZSRM, 7 satellites were selected, of which one was misclassified. 
These results suggest that under realistic classification conditions, the mode distributions tend to become more dispersed, which may affect positioning accuracy. 
Nevertheless, ZSRM consistently outperformed ZSM, despite using fewer satellites and being more susceptible to misclassification. 
This demonstrates the robustness of ZSRM and its ability to effectively refine modes through the use of GNSS reflection information. 

Looking forward, the performance of ZSRM can be further improved by enhancing the classification process itself. 
Employing deep neural network (DNN) architectures capable of capturing the complex characteristics of GNSS signals, together with more informative or domain-specific features, could strengthen the classifier’s ability to detect LOS + NLOS signals. 
Such advancements represent a promising direction for future work and may lead to additional gains in ZSRM performance beyond those demonstrated in this study.

\subsection{Processing Time Evaluation}
\label{sec:Processing}

We evaluated and compared the processing times of the ZSM and ZSRM algorithms, dividing the total computation into offline and online phases. 
In the offline phase, we measured the time required to convert building and terrain geometries---represented as vertices in a 3D city model---into constrained zonotopes. 
This phase is identical for both ZSM and ZSRM. 
The online phase, in contrast, differs between the two algorithms: ZSM includes the computation of GNSS shadows and estimation of the receiver position set, whereas ZSRM additionally incorporates GNSS reflection computation. 
Processing times were obtained by executing each algorithm ten times across all epochs and taking the average. 
All experiments were conducted on a desktop computer with an 8-core Intel Core i7-9700F CPU (3.00 GHz) and 64 GB RAM. 
The GNSS reflection computation was parallelized across reflection planes using MATLAB’s Parallel Computing Toolbox with 8-core processing, and parallel computation was also applied during GNSS shadow calculation across buildings.

The results are summarized in Table~\ref{tab:processing_time}. 
The offline phase required 172.5 s for 448 epochs. 
In the online phase, the average computation time for GNSS shadows per satellite per epoch was 0.1729 s, whereas GNSS reflections required 0.7377 s per satellite per epoch. 
A detailed breakdown of reflection computation time is provided in Table~\ref{tab:reflection_breakdown}. 
The most time-consuming component was the calculation of invisible and blocked areas, which accounted for approximately 70\% of the total reflection computation time. 
The computation time for the receiver position set per epoch was 0.0176 s for ZSM and 0.0222 s for ZSRM. 
The slightly longer duration in ZSRM results from the additional set operations associated with GNSS reflections.

\begin{table}
\centering
\caption{Comparison of processing times between ZSM and ZSRM.}
\label{tab:processing_time}
\renewcommand{\arraystretch}{1.2}
\setlength{\tabcolsep}{8pt}
\begin{tabular}{c|l|cc}
\hline
Phase & Process & ZSM & ZSRM \\
\hline
\multirow{1}{*}{Offline} 
& \begin{tabular}{@{}l@{}}
Converting buildings and terrain \vspace{-0.5mm} \\
from vertices to constrained \vspace{-0.5mm} \\ 
zonotopes (for 448 epochs)
\end{tabular}
& \multicolumn{2}{c}{172.5 s} \\
\hline
\multirow{5}{*}{Online} 
& \begin{tabular}{@{}l@{}}
Computing GNSS shadows \\ (per satellite per epoch)
\end{tabular}
& \multicolumn{2}{c}{0.1729 s} \\
\cline{2-4}
& \begin{tabular}{@{}l@{}}
Computing GNSS reflections \\ (per satellite per epoch) 
\end{tabular}
& -- & 0.7377 s \\
\cline{2-4}
& \begin{tabular}{@{}l@{}}
Estimating receiver position set \\ (per epoch) 
\end{tabular}
& 0.0176 s & 0.0222 s \\
\cline{2-4}
& \begin{tabular}{@{}l@{}}
Total (per epoch) 
\end{tabular}
& 0.1905 s & 0.9328 s \\
\hline
\end{tabular}
\end{table}

\begin{table}
\centering
\caption{Computation times for each process in GNSS reflection for ZSRM (per satellite per epoch).}
\label{tab:reflection_breakdown}
\renewcommand{\arraystretch}{1.3}
\setlength{\tabcolsep}{6pt}
\begin{tabular}{>{\raggedright\arraybackslash}p{5.0cm}|>{\centering\arraybackslash}p{2.0cm}}
\hline
\rowcolor{gray!20}
\multicolumn{2}{c}{Computing GNSS reflections (per satellite per epoch)} \\
\hline
(1) Finding reflection planes & 0.0024 s \\
\hline
(2) Computing potential areas & 0.0445 s \\
\hline
(3) Identifying invisible and blocked areas & 0.5182 s \\
\hline
(4) Calculating GNSS reflections & 0.1726 s \\
\hline
\rowcolor{gray!20}
\textbf{Total} & \textbf{0.7377 s} \\
\hline
\end{tabular}
\end{table}

The total online processing times per epoch for ZSM and ZSRM are 0.1905 s and 0.9328 s, respectively, under the assumption that satellite-level parallelization is employed. 
The current MATLAB implementation does not utilize satellite-level parallelization because MATLAB does not support nested parallelism, and parallel structures are already employed within the shadow and reflection computations for individual satellites. 
This is a MATLAB-specific limitation; an implementation in another programming language would not encounter the same issue, and satellite-level parallelization could be readily applied. 
These results indicate that both algorithms are potentially suitable for real-time position calculation with 1 Hz outputs. 
However, the absolute processing times measured in a PC and MATLAB environment are not directly meaningful, since an actual implementation on an embedded receiver board would differ substantially. 
Nevertheless, ZSRM is approximately five times slower than ZSM. 
Therefore, a receiver with sufficient computational capacity could implement the ZSRM algorithm to achieve improved positioning accuracy.

\section{Conclusion} 
\label{sec:Conclusion}

In this paper, we proposed the ZSRM algorithm, a 3DMA GNSS technique designed to enhance positioning accuracy in urban areas. 
In this approach, objects such as 3D buildings are represented as sets using constrained zonotopes. 
ZSRM leverages these constrained zonotopes to compute GNSS shadows and reflections. 
By refining the coarse set-based receiver position with both GNSS shadows and reflections, ZSRM achieves higher positioning performance than the existing ZSM technique, which relies solely on GNSS shadows.
We evaluated the performance of ZSRM using GNSS data collected in a real urban environment. 
The results demonstrated an improvement in RMS horizontal position error ranging from 10.0\% to 53.6\% compared with ZSM. 
Furthermore, the RMS position bounds improved by 18.0\% to 50.1\% in the cross-street direction and by 30.7\% to 59.3\% in the along-street direction. 

While ZSRM provides significant performance gains, its hard-decision strategy based on signal classification can be affected by errors from misclassified satellites.
In future work, we plan to explore probabilistic frameworks to further improve robustness.
Additionally, incorporating low Earth orbit (LEO) satellite signals \cite{Neinavaie23} could further enhance ZSRM’s robustness by increasing satellite availability and geometric diversity in challenging urban environments.
Moreover, leveraging terrestrial 5G ranging signals \cite{Zhang24:A, Zhang25:Cooperative} could improve the positioning precision of ZSRM in dense urban environments.
Overall, ZSRM represents a promising approach for enhancing urban GNSS accuracy by jointly exploiting shadows and reflections, offering significant improvements over conventional methods.

\section*{Acknowledgment}

The authors acknowledge the use of a generative AI tool (ChatGPT, OpenAI) to assist with grammar and language editing during manuscript preparation.  
The AI tool was not used to generate technical content, ideas, data, or citations.  
All technical content, methodology, analysis, and conclusions are the sole work of the authors.

\bibliographystyle{IEEEtran}
\bibliography{mybibfile, IUS_publications}

\begin{IEEEbiography}[{\includegraphics[width=1in,height=1.25in,clip,keepaspectratio]{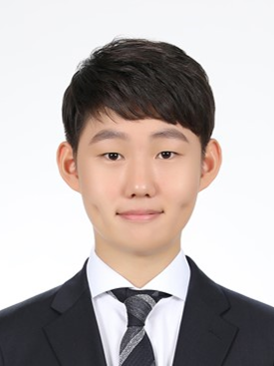}}]{Sanghyun Kim} is an M.S./Ph.D. student with the School of Integrated Technology, Yonsei University, Incheon, Republic of Korea. 
He received the B.S. degree in Integrated Technology from Yonsei University. 
His research interests include seamless positioning in urban environments and intelligent transportation systems. 
Mr. Kim received undergraduate and graduate fellowships from the Information and Communications Technology (ICT) Consilience Creative Program, supported by the Ministry of Science and ICT, Republic of Korea. 
\end{IEEEbiography}

\begin{IEEEbiography}[{\includegraphics[width=1in,height=1.25in,clip,keepaspectratio]{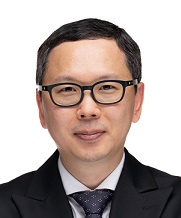}}]{Jiwon Seo} (Member, IEEE) received the B.S. degree in mechanical engineering (division of aerospace engineering) from the Korea Advanced Institute of Science and Technology (KAIST), Daejeon, Republic of Korea, in 2002. 
He earned the M.S. degrees in aeronautics and astronautics (2004) and in electrical engineering (2008), as well as the Ph.D. degree in aeronautics and astronautics (2010), all from Stanford University, Stanford, CA, USA. 
He is currently an Underwood Distinguished Professor at Yonsei University, Incheon, Republic of Korea, where he is a Professor and Chair with the School of Integrated Technology. 
He also serves as an Adjunct Professor with the Department of Convergence IT Engineering at Pohang University of Science and Technology (POSTECH), Pohang, Republic of Korea. 
His research interests include GNSS anti-jamming technologies, complementary PNT systems, and intelligent unmanned systems. 
Dr. Seo is a member of the International Advisory Council of the Resilient Navigation and Timing Foundation, Alexandria, VA, USA, and of the Advisory Committee on Defense of the Presidential Advisory Council on Science and Technology, Republic of Korea.
\end{IEEEbiography}

\vfill

\end{document}